\documentclass[aps,prb,footinbib,showpacs,twocolumn,showkeys,amssymb,amsmath,superscriptaddress,groupedaddress]{revtex4-1}
\pdfoutput=1

\usepackage{hyperref}

\usepackage{amsmath}
\usepackage{braket}
\usepackage{graphicx}
\usepackage{graphics}
\usepackage{color}
\usepackage{stackengine}

\usepackage{verbatim}

\begin{document}

%\title{Long-range AAH model: exact prescription for the fraction of delocalized eigenstates}
%\title{Long-range quasicrystals: prescription for the fraction of delocalized eigenstates}
\title{Fraction of delocalized eigenstates in the long-range AAH model}

\author{Nilanjan Roy}
\author{Auditya Sharma}
\affiliation{Department of Physics, Indian Institute of Science Education and Research, Bhopal, Madhya Pradesh 462066, India}

\date{\today}

\begin{abstract}
We uncover a systematic structure in the single particle phase-diagram
of the quasiperiodic Aubry-Andr\'e-Harper(AAH) model with power-law
hoppings ($\sim \frac{1}{r^\sigma}$) when the quasiperiodicity
parameter is chosen to be a member of the `metallic mean family' of
irrational Diophantine numbers. In addition to the fully delocalized
and localized phases we find a co-existence of multifractal
(localized) states with the delocalized states for $\sigma<1$
($\sigma>1$). The fraction of delocalized eigenstates in these phases
can be obtained from a general sequence, which is a manifestation of a
mathematical property of the `metallic mean family'. The entanglement
entropy of the noninteracting many-body ground states respects the
area-law if the Fermi level belongs in the localized regime while
logarithmically violating it if the Fermi-level belongs in the
delocalized or multifractal regimes. 
The prefactor of logarithmically violating term shows interesting behavior in different phases. Entanglement entropy shows the
area-law even in the delocalized regime for special filling fractions,
which are related to the metallic means.
\end{abstract}

\maketitle

\section{Introduction}\label{lrh_intro}
Quasiperiodic systems or quasicrystals lie at the junction of periodic
and random systems and exhibit non-trivial intermediate localization
properties~\cite{goldman1993quasicrystals,kohmoto_rev,albuquerque2003theory,macia2005role}.
Unlike the one-dimensional Anderson model~\cite{anderson} where even
an infinitesimal random potential leads to localization, a non-zero
finite quasiperiodic potential is essential for the
Aubry-Andr\'e-Harper(AAH) model to show a delocalization-localization
transition in one dimension~\cite{aubry,harper}. Remarkably, even the
presence of a mobility edge (which in the traditional Anderson model
can be seen only in three dimensions) has been reported in variants of
the AAH model~\cite{sarma1990localization,ganeshan2015nearest} even in
one dimension. The AAH potential has been realized in experiments of
ultracold atoms studying single particle
localization~\cite{lahini,lucioni2011observation,lye2005bose} and
`many body localization'~\cite{bordia}, which has lead to a fresh wave
of interest in quasiperiodic systems at zero
~\cite{oganesyan2007localization,alet,abanin2019colloquium,iyer,modak2015many,vznidarivc2018interaction,xu2019butterfly}
and finite
temperatures~\cite{nessi2011finite,jstat,michal2014delocalization}
in recent times. On the other hand, the study of Hamiltonians with power-law hoppings or interactions
$(\propto\frac{1}{r^\sigma})$ has seen a resurgence of interest after
such Hamiltonians were realized in experiments of ultra-cold
systems~\cite{Kim2,gorshkov2014,Britton,rajibul,Robert,Weimer1,Labuhn,Schauss1455,baier2018realization,Alexey,Manmana,Hazzard}. When
the hopping strength is sufficiently long-ranged, instead of the
exponentially localized eigenstates seen in short-range models, one
may obtain algebraically localized
eigenstates~\cite{lima1,mirlin1,celardo2016shielding,LSantos}. Despite the rich literature on 
both quasiperiodic and long-range systems, the interplay of both these aspects
has only begun to be studied~\cite{deng2019one,gopalakrishnan2017self,saha2019anomalous,modak2020many}.

The effect of power-law hoppings which breaks the self-duality of the
quasi-periodic AAH potential has been studied very
recently~\cite{deng2019one}.  This study has shown the appearance of
multifractal (localized) eigenstates which co-exist with delocalized
eigenstates for $\sigma<1$ $(\sigma>1)$~\cite{deng2019one}. The
irrationality of the quasiperiodicity parameter ($\alpha$) which
renders the Hamiltonian quasiperiodic, is key to the striking physics
of this system. In the present work we show, with the aid of a
transparent prescription, the relationship between the fraction of
delocalized eigenstates in the different phases of the system and the
parameter $\alpha$. While most studies of the AAH model choose this
irrational number to be the golden mean ($(\sqrt{5}-1)/2$), we obtain
a general result for a broader class of irrrational Diophantine
numbers referred to as the `the metallic mean family', of which the
golden mean is just one element.
\begin{figure}
\centering
	\includegraphics[width=7.0cm,height=4.9cm]{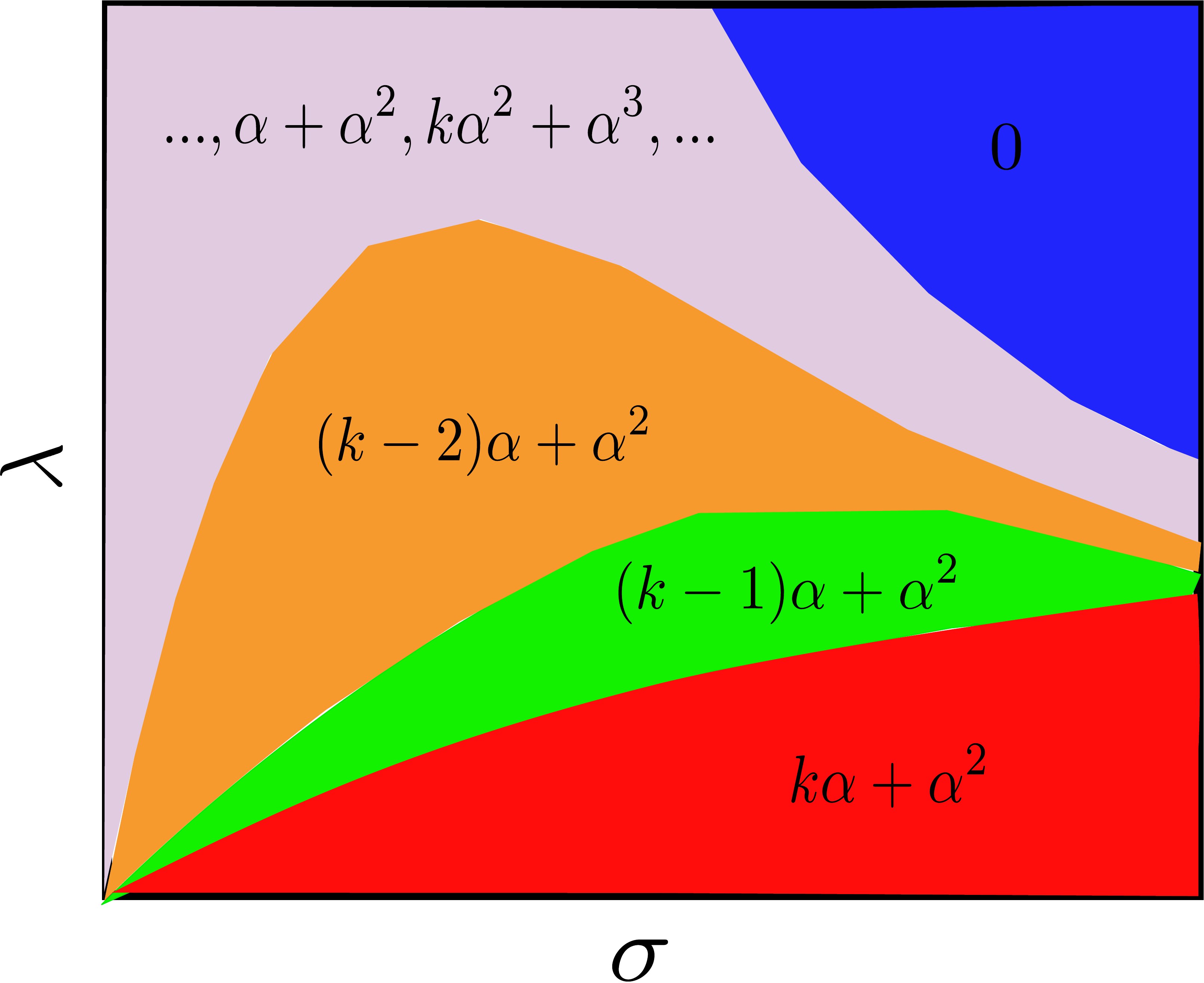}
	\caption{Schematic of the phases of a single particle in the
          LRH model for the quasiperiodicity parameter $\alpha$,
          shown in different colors. The colored phases are also
          labelled by the fraction of delocalized eigenstates ($\eta$)
          as shown in the figure. Here $k=1,2,3$ when $\alpha$ is
          `golden mean', `silver mean' and `bronze mean'
          respectively. The strength of the quasiperiodic potential
          and power-law hopping parameter are denoted as $\lambda$ and
          $\sigma$ respectively.}
	\label{schematic_lrh}
\end{figure}  

In this work, we chart out the phase diagram of a single particle in
the presence of the AAH potential and power-law hoppings when $\alpha$
is set to be a member of the `metallic mean family', with special
attention given to the `golden mean', `silver mean' and `bronze
mean'. In addition to the delocalized and localized phases, we obtain
mixed phases where the multifractal (localized) states can co-exist
with delocalized states for $\sigma<1$ ($\sigma>1$). One of the key
findings of our work is that the fraction of delocalized eigenstates
in these phases can be obtained from a general sequence, which is
related to a mathematical property of the metallic means (see
schematic in Fig.~\ref{schematic_lrh}).  Moreover we study the subsystem size scaling of
entanglement entropy~\cite{eisert,laflorencie2016quantum} 
of the noninteracting fermionic many-body ground states to
characterize different phases in the model. The delocalized and multifractal Fermi level shows logarithmic violation of the area-law of entanglement entropy with different prefactors in the logarithm term. The prefactor is found to vary in different phases.
In the delocalized regime entanglement entropy is surprisingly found to follow the area-law for
the special filling fractions which are related to the metallic
means. We show that such strange behavior at the special filling fractions~\cite{roy2019study}, may be
understood from the single particle spectrum.

The paper is organized as follows. In Sec.~\ref{lrh_sec1} we describe the model and metallic means. In Sec.~\ref{lrh_sec2} we derive the single particle phase diagram for metallic means by analyzing fractal dimension and inverse participation ratio of the eigenstates. In Sec.~\ref{lrh_sec3} we show a general sequence to obtain the fraction of delocalized states in different phases. In Sec.~\ref{lrh_sec4} we analyze the scaling of the ground state entanglement entropy of non-interacting fermions to characterize the phases. Then we conclude in Sec.~\ref{lrh_sec5}. 

\section{The model}\label{lrh_sec1}
The model of interest is the one dimensional long-range
AAH (LRH) model given by the Hamiltonian:
\begin{eqnarray}
H = -\sum\limits_{i<j}^{N} \bigg(\frac{J}{r_{ij}^\sigma} \hat{c}_i^\dagger \hat{c}_j + H.c.\bigg)\nonumber + \lambda\sum\limits_{i=1}^{N} \cos(2\pi\alpha i + \theta_p)\hat{n}_i,\nonumber\\
\label{ham}
\end{eqnarray}
where $\hat{c}_i^\dagger$ $(\hat{c}_i)$ represents the single particle
creation (destruction) operator at site $i$ and corresponding number operator $\hat{n}_i=\hat{c}_i^\dagger \hat{c}_i$. We consider a lattice of total number of sites $N$, where
$r_{ij}$ is the geometric distance between the sites $i$ and $j$ in a
ring. Here $\lambda$ is the strength of the quasiperiodic potential
with the parameter $\alpha$ chosen to be a Diophantine irrational
number~\cite{modugno2009exponential} e.g. $(\sqrt{5}-1)/2$. $\theta_p$
is an arbitrary global phase. The strength of the long range hopping
is controlled by $J$ and the long range parameter in the hopping
$\sigma$. We will assume $J=1$ for all the numerics. In the
$\sigma\to\infty$ limit this model becomes the celebrated
Aubry-Andr\'e-Harper (AAH) model~\cite{aubry,harper}. As a consequence
of self-duality~\cite{aubry,harper}, all the eigenstates are
delocalized for $\lambda<2J$ and localized for
$\lambda>2J$~\cite{modugno2009exponential}. For a finite $\sigma$ self-duality
is broken.

{\it Metallic mean family}:  Any irrational number can be written as a continued
fraction\cite{continuedfraction} which allows for a successive rational
approximation of it in the form of $a/b$ where $a$, $b$ are co-prime
numbers. For Diophantine numbers there always
exists a lower bound to how closely such irrational numbers may be represented by rational approximations, such that $|\alpha-\frac{a}{b}|>\epsilon/b^{2+\zeta}$ with $\epsilon>0$ and $\zeta\geq 0$
\cite{diophantine,serda2015classificaction}. The above property is a sign of the
strength of the irrationality of Diophantine numbers. 

It is useful to consider  a generalized
`$k$-Fibonacci sequence'~\cite{fibonacci}, given by
\begin{eqnarray}
F_u = k F_{u-1} + F_{u-2},
\label{fibo}
\end{eqnarray} 
with $F_0=0$, $F_1=1$. The limit $\alpha = \lim_{u\rightarrow\infty}
F_{u-1}/F_{u}$ with $k=1,2,3...$ yields the `metallic mean family',
the first three members of which are the well-known `golden mean'
($\alpha_g=(\sqrt{5}-1)/2$), the `silver mean' ($\alpha_s=\sqrt{2}-1$)
and `bronze mean' ($\alpha_b=(\sqrt{13}-3)/2$) respectively.  A slowly
converging sequence of rational approximations of these Diophantine
numbers is given by $F_{u-1}/F_{u}$ for two successive members in the
sequence for a fixed integer $k$. Each member $\alpha$ of the
`metallic mean' family satisfies the following relation:
\begin{eqnarray}
(\alpha)^{z} = k (\alpha)^{z+1} + (\alpha)^{z+2},
\label{fraction}
\end{eqnarray}  
where $k=1,2,3..$ for $\alpha=\alpha_g,\alpha_s,\alpha_b,..$
respectively, and $z$ is a non-negative integer. Putting $z=0$ in
Eq.~\ref{fraction} also yields an important case namely, $k \alpha +
\alpha^2 = 1$.
\begin{figure*}
	\centering
	        \stackunder{\includegraphics[width=0.63\columnwidth,height=4.3cm]{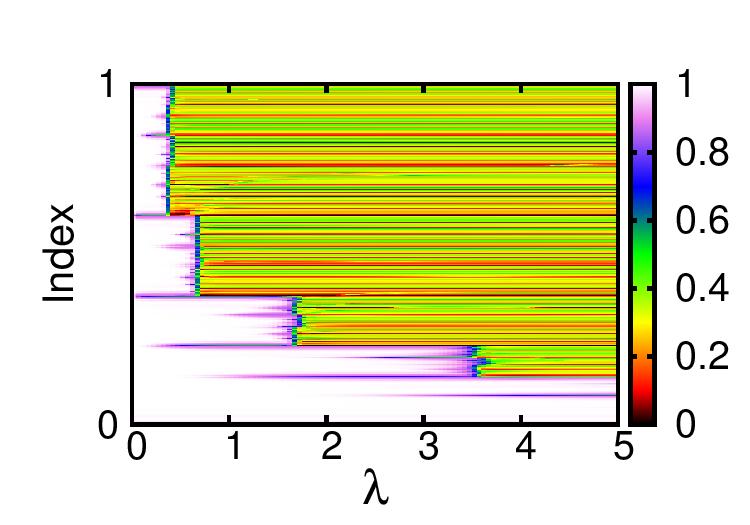}}{(a)}
		  	\stackunder{\includegraphics[width=0.63\columnwidth,height=4.3cm]{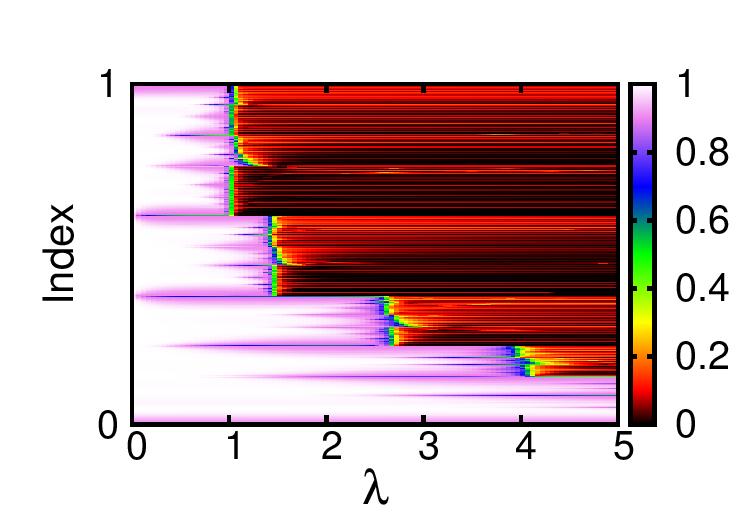}}{(b)}
		  	\stackunder{\includegraphics[width=0.63\columnwidth,height=4.3cm]{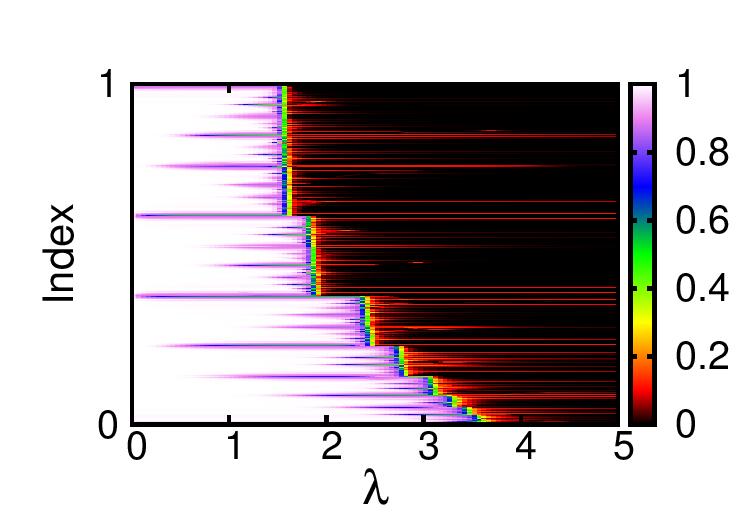}}{(c)}
		  	\stackunder{\includegraphics[width=0.63\columnwidth,height=4.3cm]{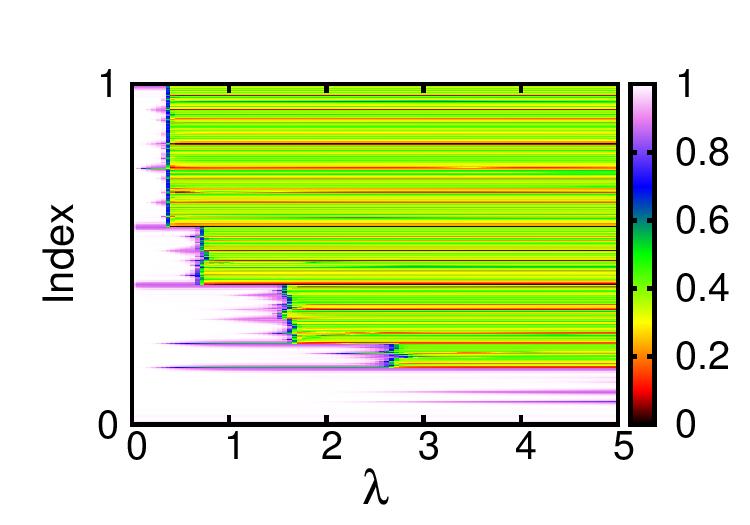}}{(d)}
		  	 \stackunder{\includegraphics[width=0.63\columnwidth,height=4.3cm]{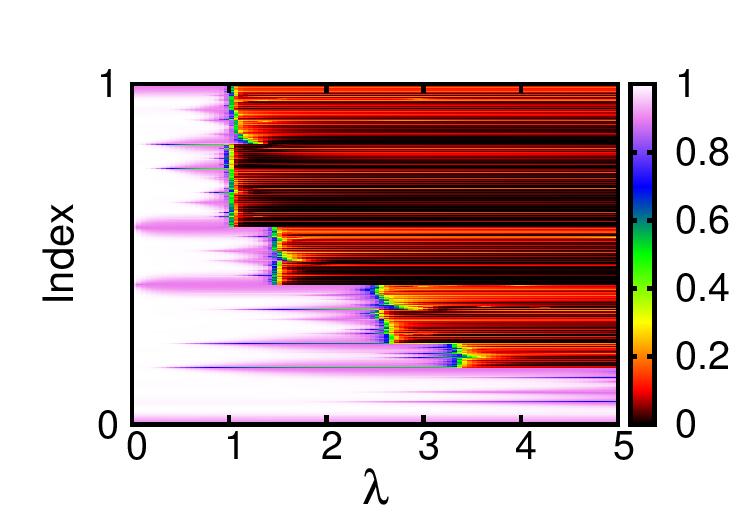}}{(e)}
		  	 \stackunder{\includegraphics[width=0.63\columnwidth,height=4.3cm]{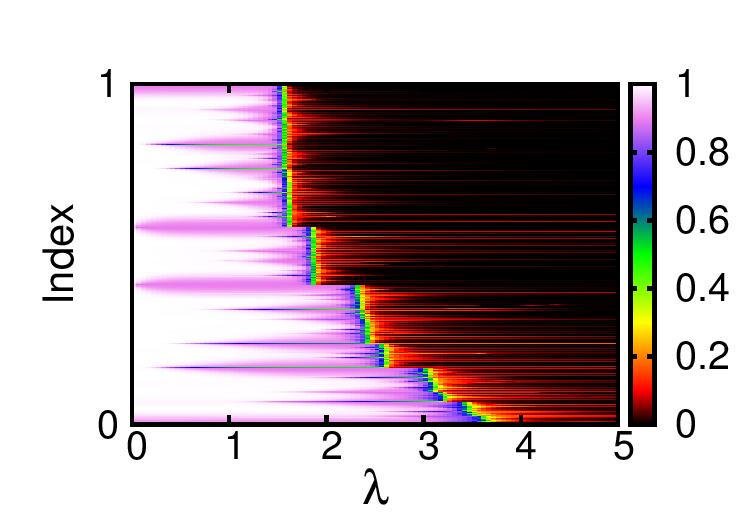}}{(f)}
		  	 \stackunder{\includegraphics[width=0.63\columnwidth,height=4.3cm]{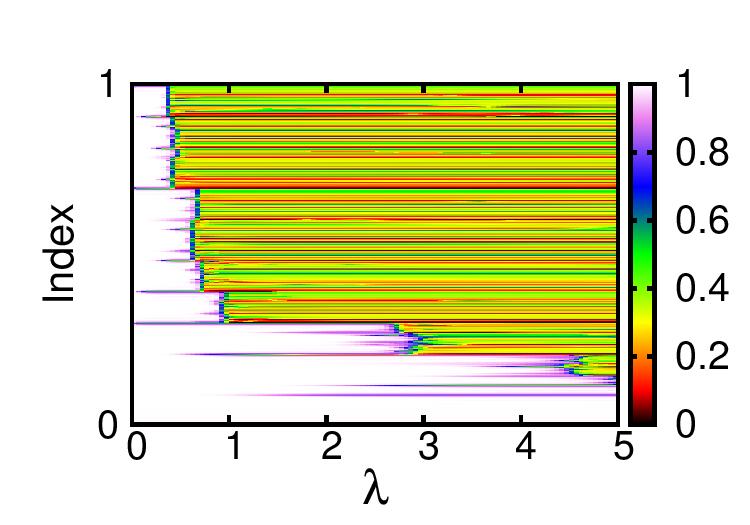}}{(g)}
		  	 \stackunder{\includegraphics[width=0.63\columnwidth,height=4.3cm]{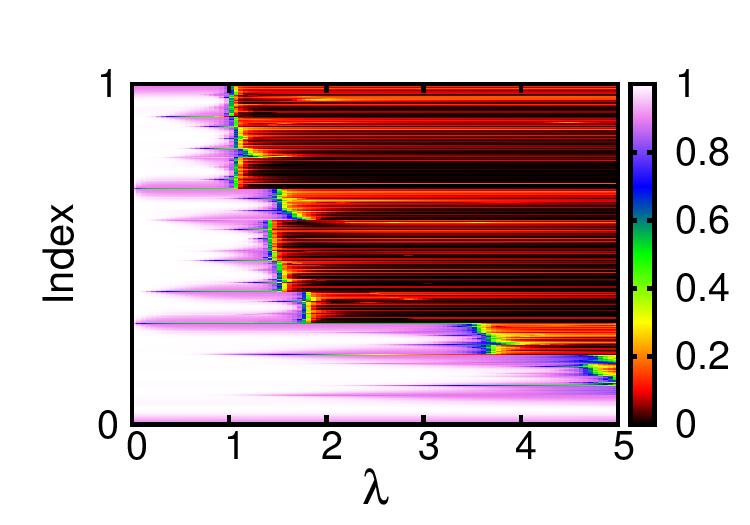}}{(h)}
		  	 \stackunder{\includegraphics[width=0.63\columnwidth,height=4.3cm]{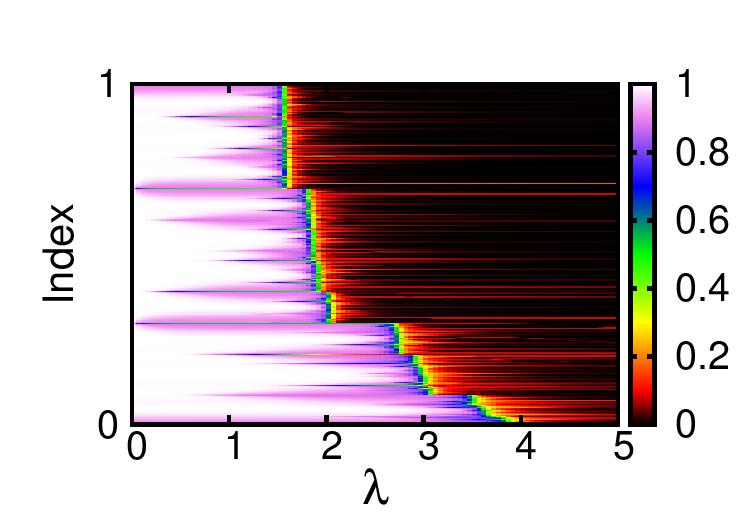}}{(k)}
	\caption{(a-c) Fractal dimension $D_2$ (in color) as a
          function of $\lambda$ and increasing fractional eigenstate
          index $n/N$ starting from the ground state for $\alpha_g$
          and $\sigma=0.5, 1.5$ and $3.0$ respectively. 
          (d-f) Same plots for $\alpha_s$.
          (g-k) Same plots for $\alpha_b$.
          For all the
          plots, $N=1000$ and $\delta=0.02$.}
	\label{D2_lrh}
\end{figure*}

\section{Single particle properties}\label{lrh_sec2}
Now we consider a single particle in the LRH
model with different parameters $\alpha_g,\alpha_s$ and
$\alpha_b$ which are members of the `metallic mean family'. In order to determine the phases we calculate the fractal dimension and inverse participation ratio of the eigenstates for
$\theta_p=0$. 

\subsection{Fractal dimension}
We employ the box
counting procedure to determine the fractal
dimension~\cite{jensen,janssen1994mutifractal,huckenstein,ECuevas}.
Dividing the system of $N$ sites into $N_l=N/l$ boxes of $l$ sites
each, the `fractal dimension' is defined as:
\begin{eqnarray}
D_f = \lim\limits_{\delta\rightarrow 0}\frac{1}{f-1}\frac{\ln \sum_{m=1}^{N_l} {(\mathcal I_m)}^f}{\ln \delta},
\end{eqnarray}
where $\mathcal I_m=\sum_{i\in m}|\psi_n(i)|^{2}$ computed inside the
$m\textsuperscript{th}$ box for the $n\textsuperscript{th}$ eigenstate
$\ket{\psi_n}$ and $\delta=1/N_l$. In the perfectly delocalized
(localized) phase $D_f$ is unity (zero), whereas for a multifractal
state $D_f$ shows a non-trivial dependence on $f$ and $0<D_f<1$.

\begin{figure}
	\centering
	        \stackunder{\includegraphics[width=4.0cm,height=3.7cm]{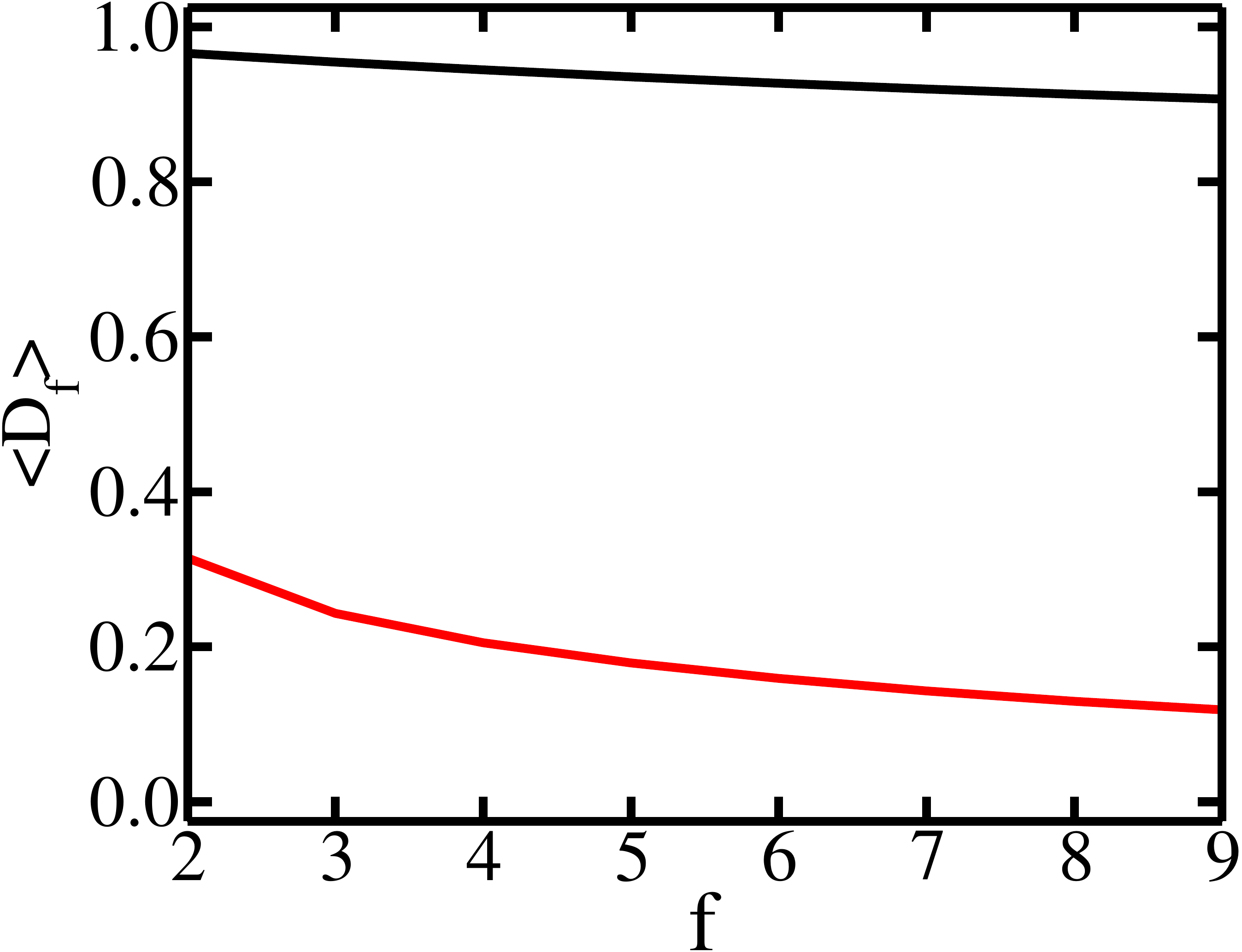}}{(a)}\hspace{0.15cm}
		  	\stackunder{\includegraphics[width=4.0cm,height=3.7cm]{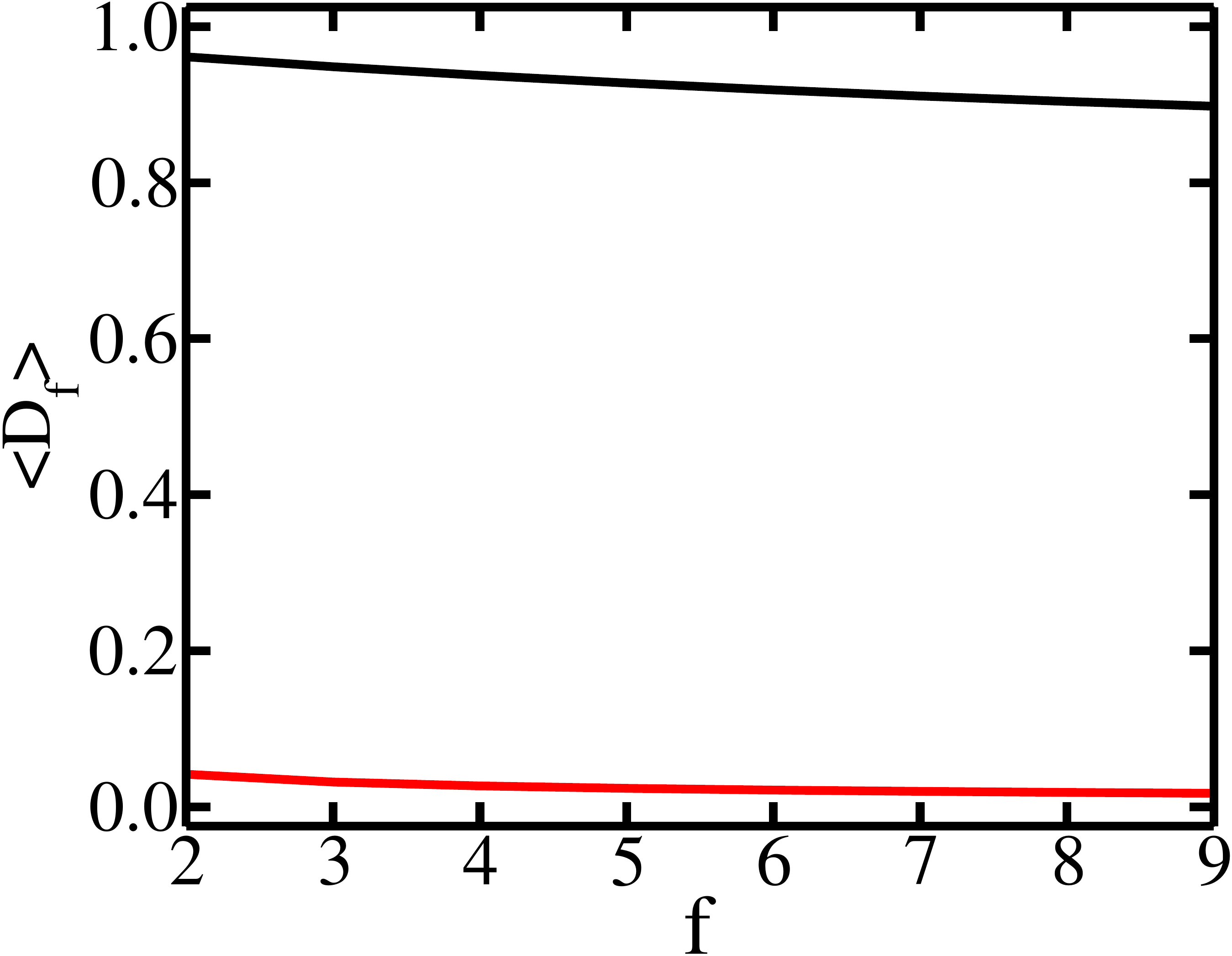}}{(b)}
		  	\caption{(a) Averaged $\langle D_f\rangle$ as a function of
                          $f$ for $\lambda=1.0$ and $\sigma=0.5$ for
                          which the system is in the $P_2$ phase with
                          a DM edge. (b) Similar plots for
                          $\lambda=2.0$ and $\sigma=1.5$ for which the
                          system is in the $P_2$ phase with a DL
                          edge. $\langle D_f\rangle$ is calculated by averaging
                          over $\alpha_g^2$ fraction of delocalized
                          and $(1-\alpha_g^2)$ fraction of
                          multifractal/localized eigenstates. For all
                          the plots system size $N=987$ and
                          $\delta=1/N_l=0.02$.}
	\label{Dq_lrh}
\end{figure}

Fig.~\ref{D2_lrh}(a-c) shows $D_2$ as a function of $\lambda$ for
all the single particle eigenstates when the quasiperiodicity
parameter is fixed at $\alpha_g$ for $\sigma=0.5,1.5$ and $3.0$
respectively. As can be seen from Fig.~\ref{D2_lrh}(a) for
$\sigma=0.5$, the fraction of delocalized eigenstates decreases and
fractal states ($0<D_2<1$) appear in blocks as $\lambda$ increases.
It turns out that these states are actually multifractal which we discuss later (See Fig.~\ref{Dq_lrh}).
Hence there exists a
delocalized-to-multifractal (DM) edge in the eigenstate spectrum. The
DM edge goes down in steps as the fraction of delocalized eigenstates
decreases with $\lambda$. However, the position of the DM edge remains
unchanged within each step as the fraction of delocalized eigenstates
(denoted as $\eta$ hereafter) stays constant in that region. It is
found that in the decreasing step-like regions defined by constant DM
edges, $\eta=\alpha_g,\alpha_g^2,\alpha_g^3,...$.  We denote the
step-like regions as $P_q$ $(q=1,2,3,...)$ phases with
$\eta=\alpha_g,\alpha_g^2,\alpha_g^3,...$
respectively.
\begin{figure*}
\centering
	\stackunder{\includegraphics[width=5.2cm,height=4.1cm]{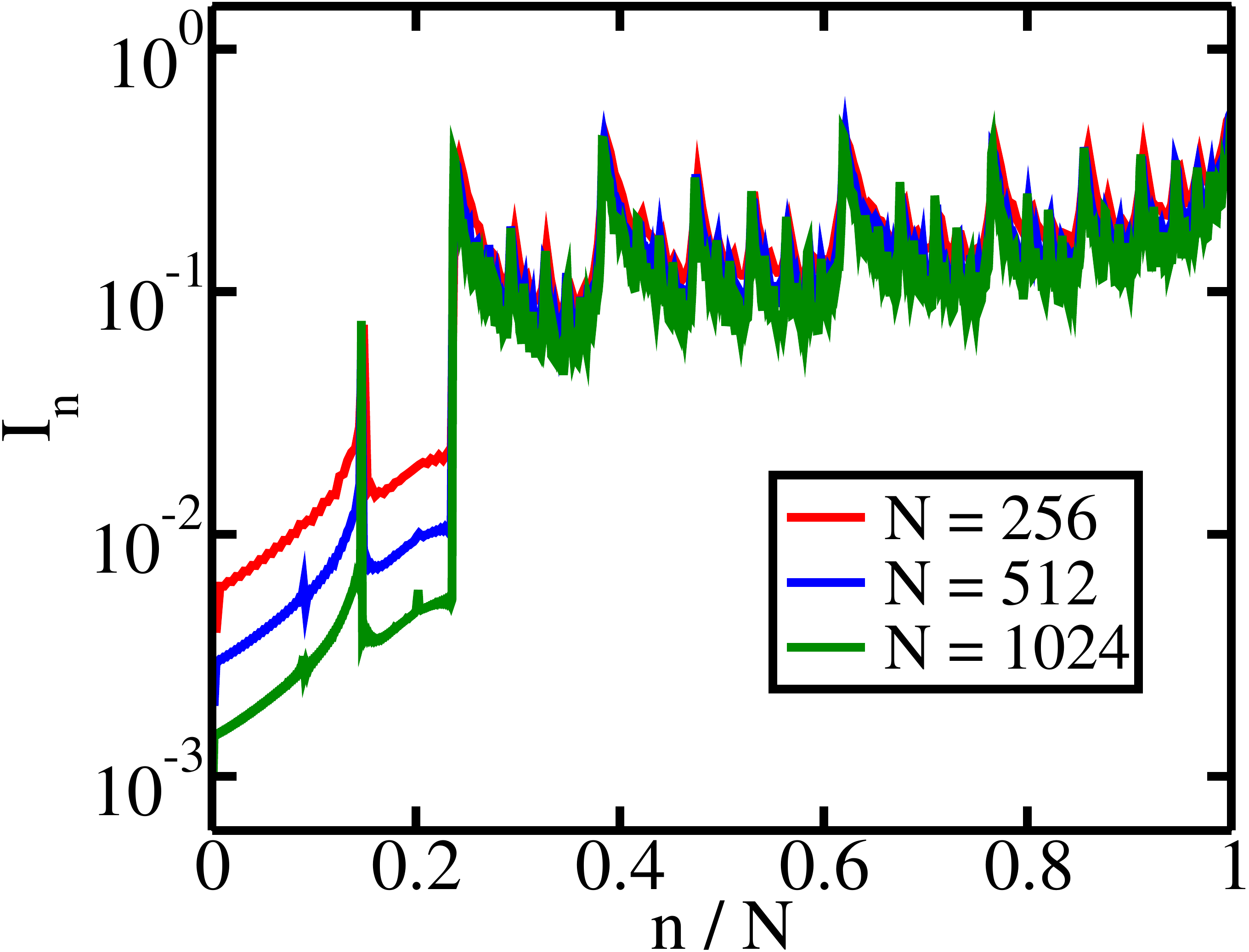}}{(a)}
	\stackunder{\includegraphics[width=5.2cm,height=4.1cm]{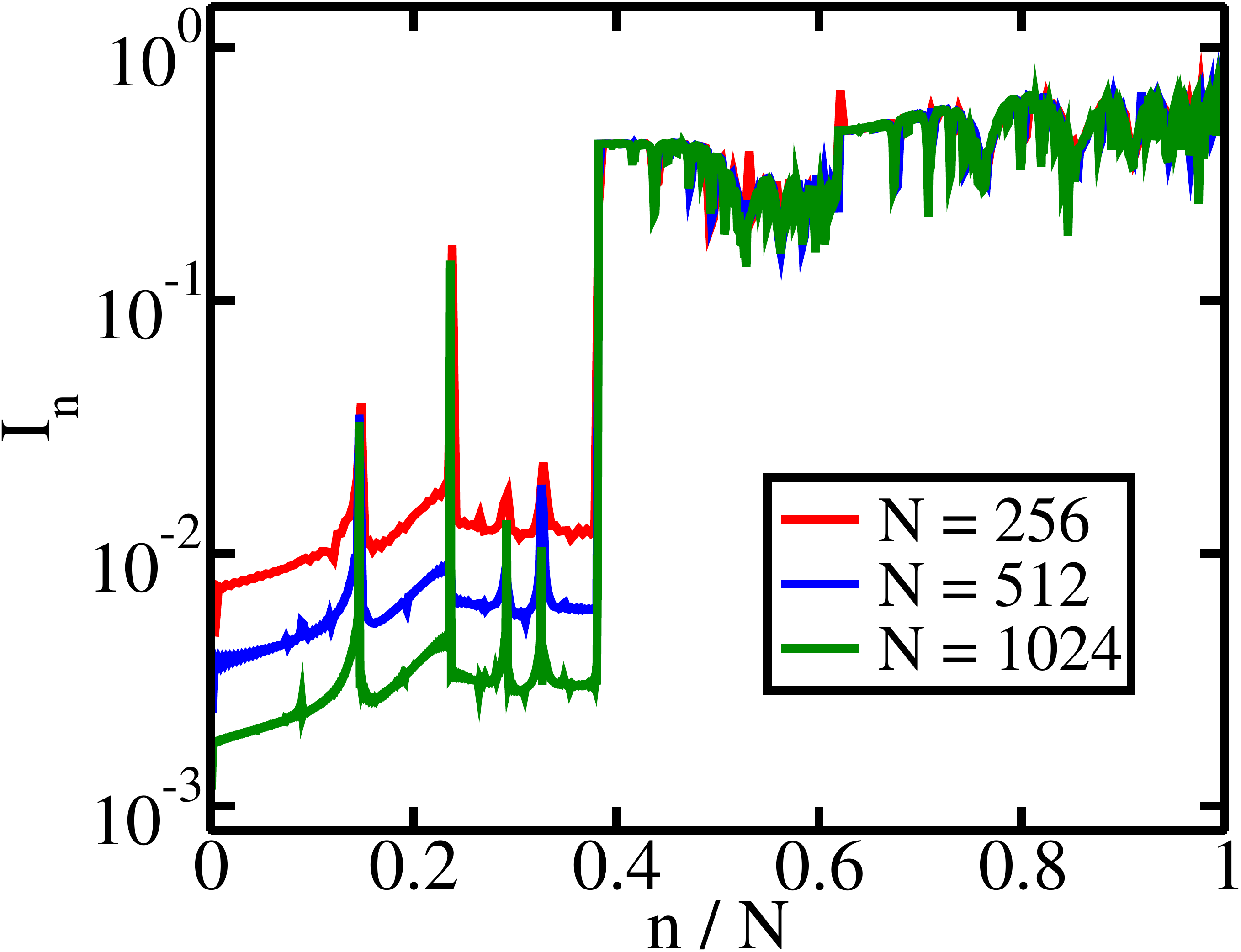}}{(b)}
	\stackunder{\includegraphics[width=5.2cm,height=4.1cm]{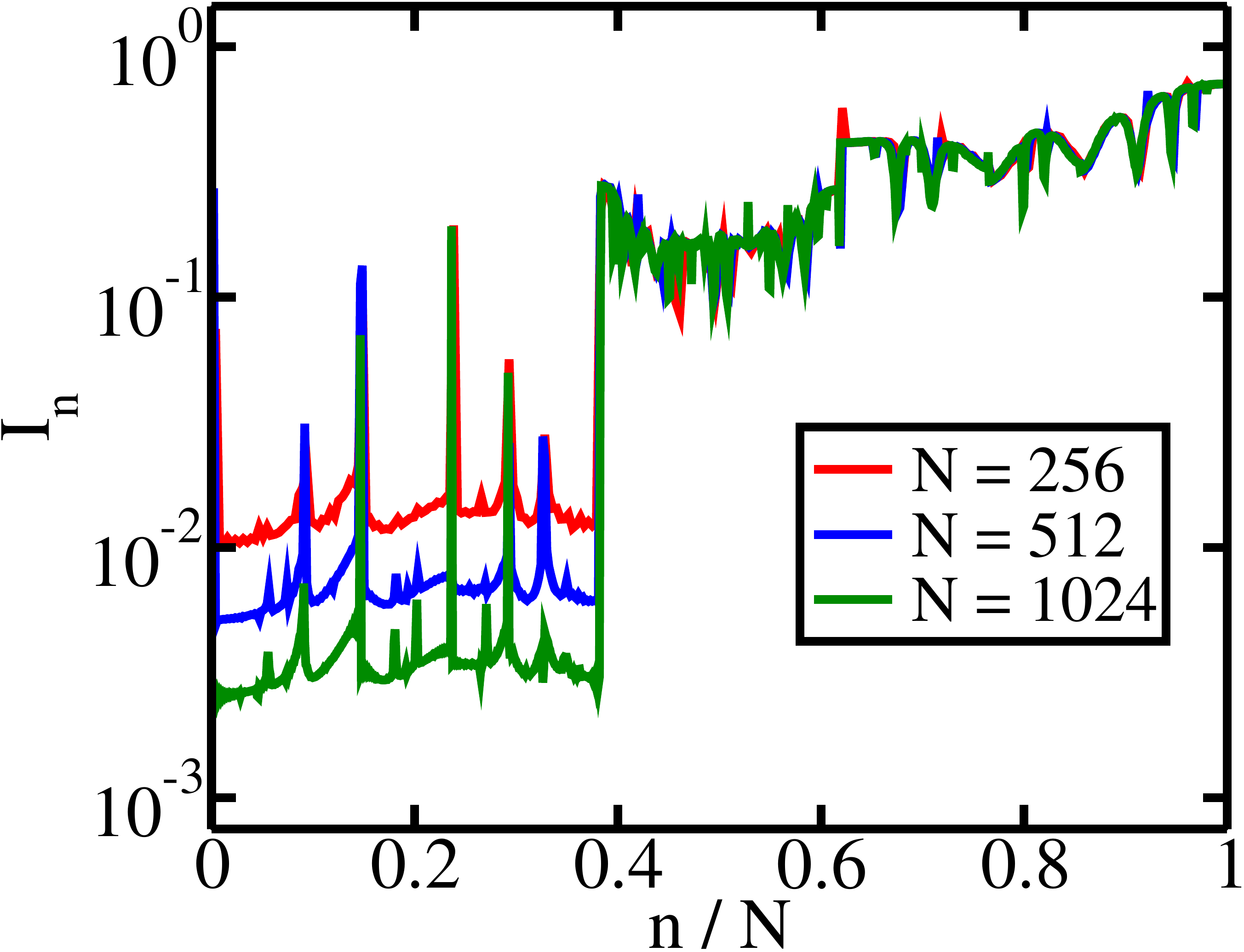}}{(c)} 
	\stackunder{\includegraphics[width=5.2cm,height=4.1cm]{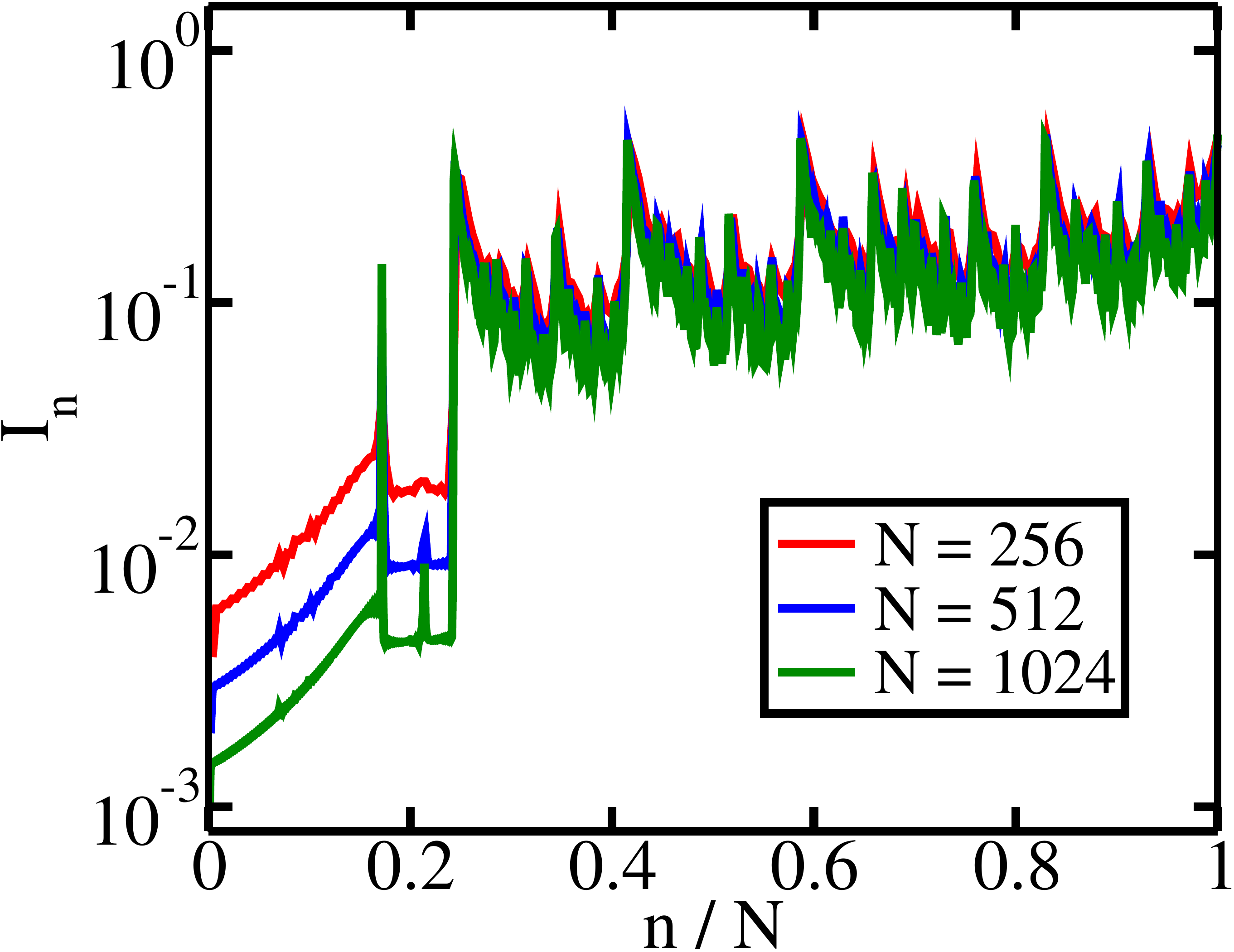}}{(d)}
	\stackunder{\includegraphics[width=5.2cm,height=4.1cm]{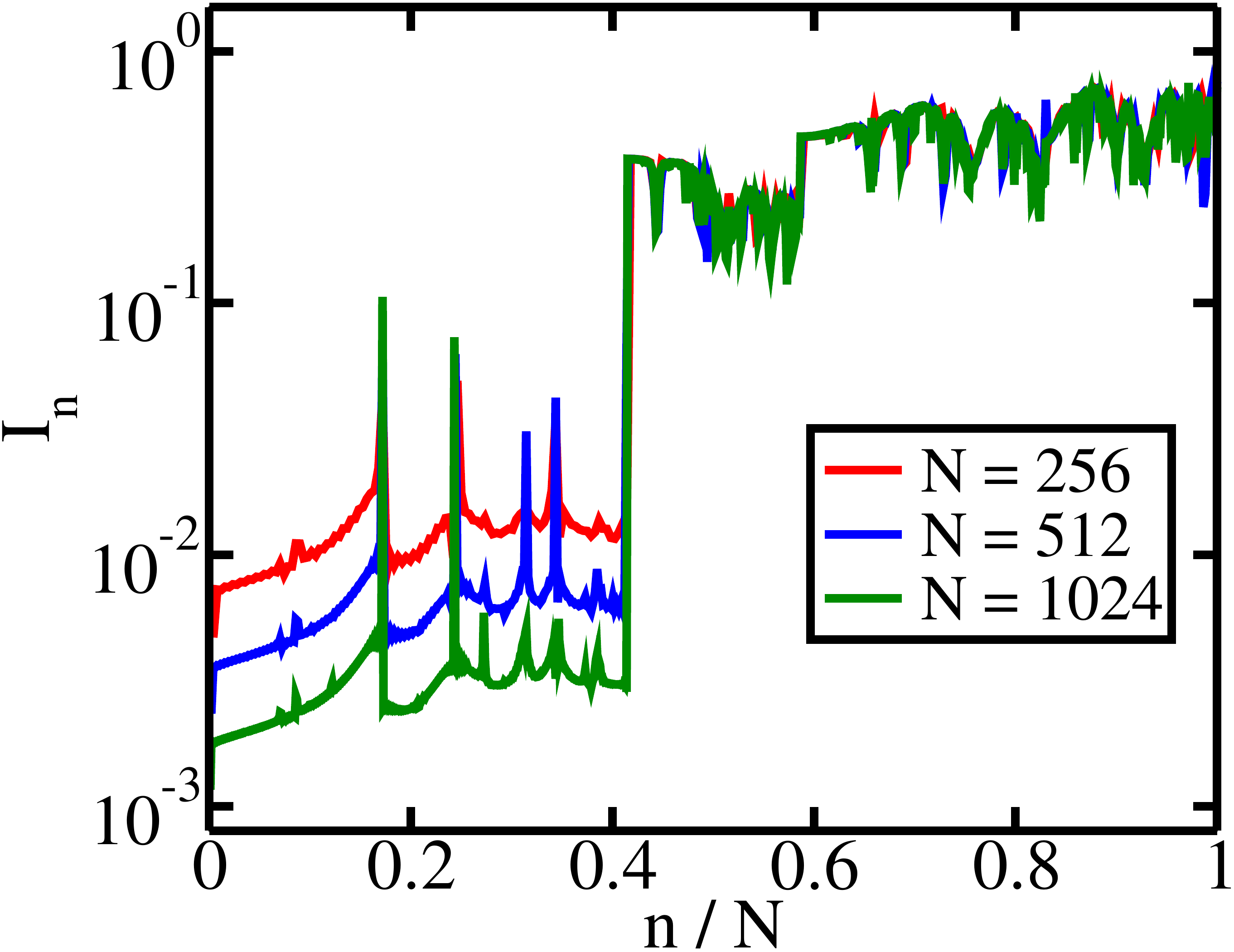}}{(e)}
	\stackunder{\includegraphics[width=5.2cm,height=4.1cm]{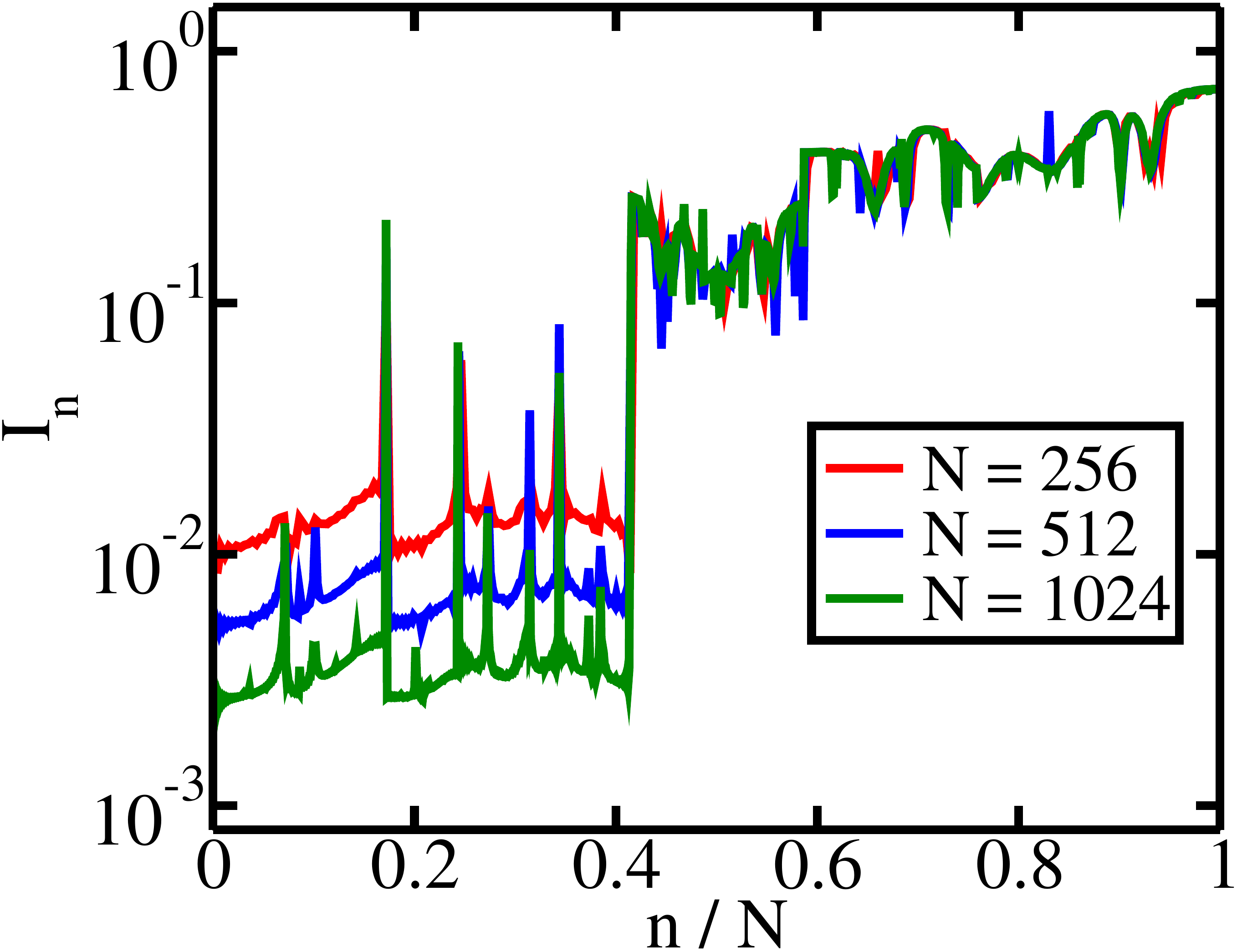}}{(f)} 
	\stackunder{\includegraphics[width=5.2cm,height=4.1cm]{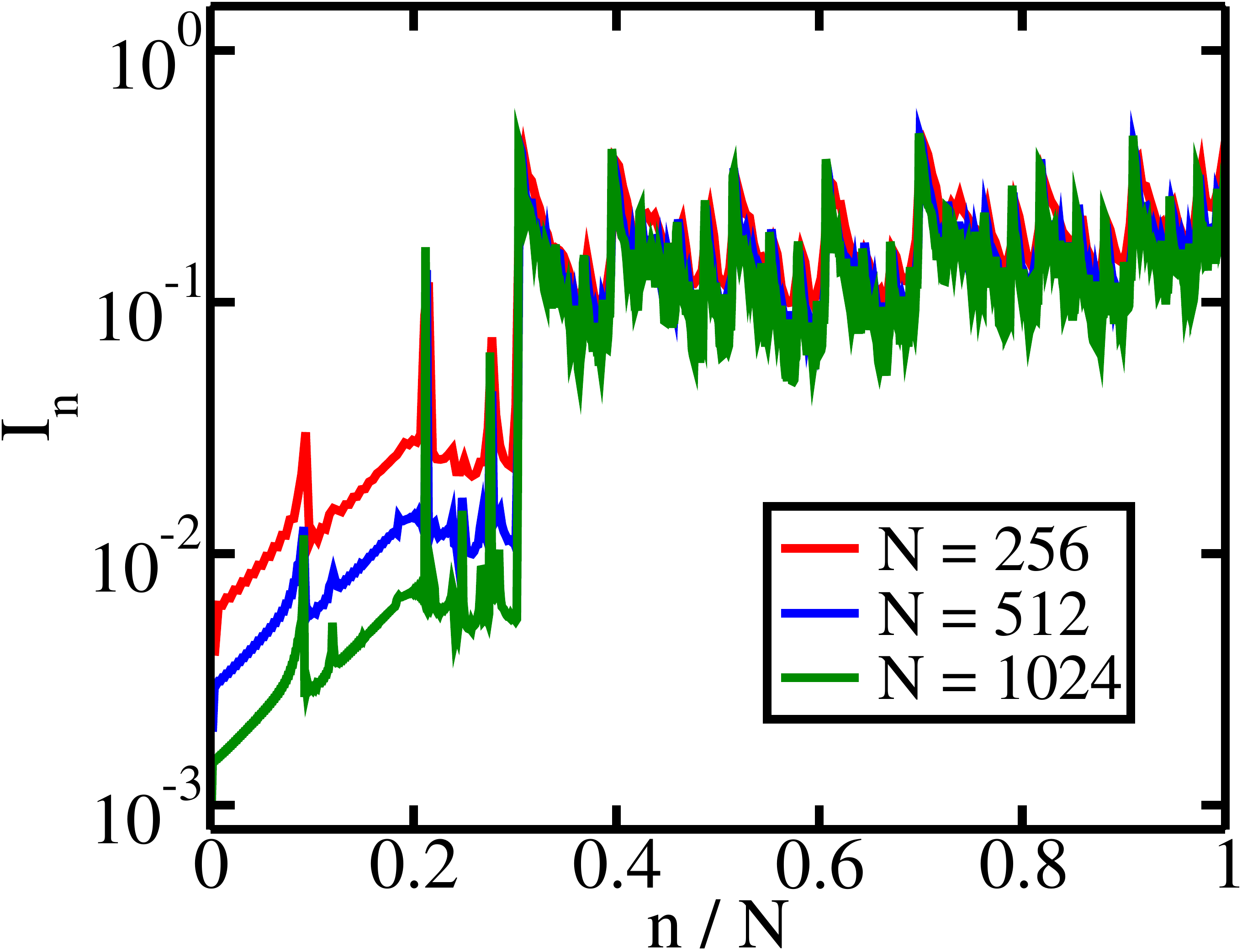}}{(g)}
	\stackunder{\includegraphics[width=5.2cm,height=4.1cm]{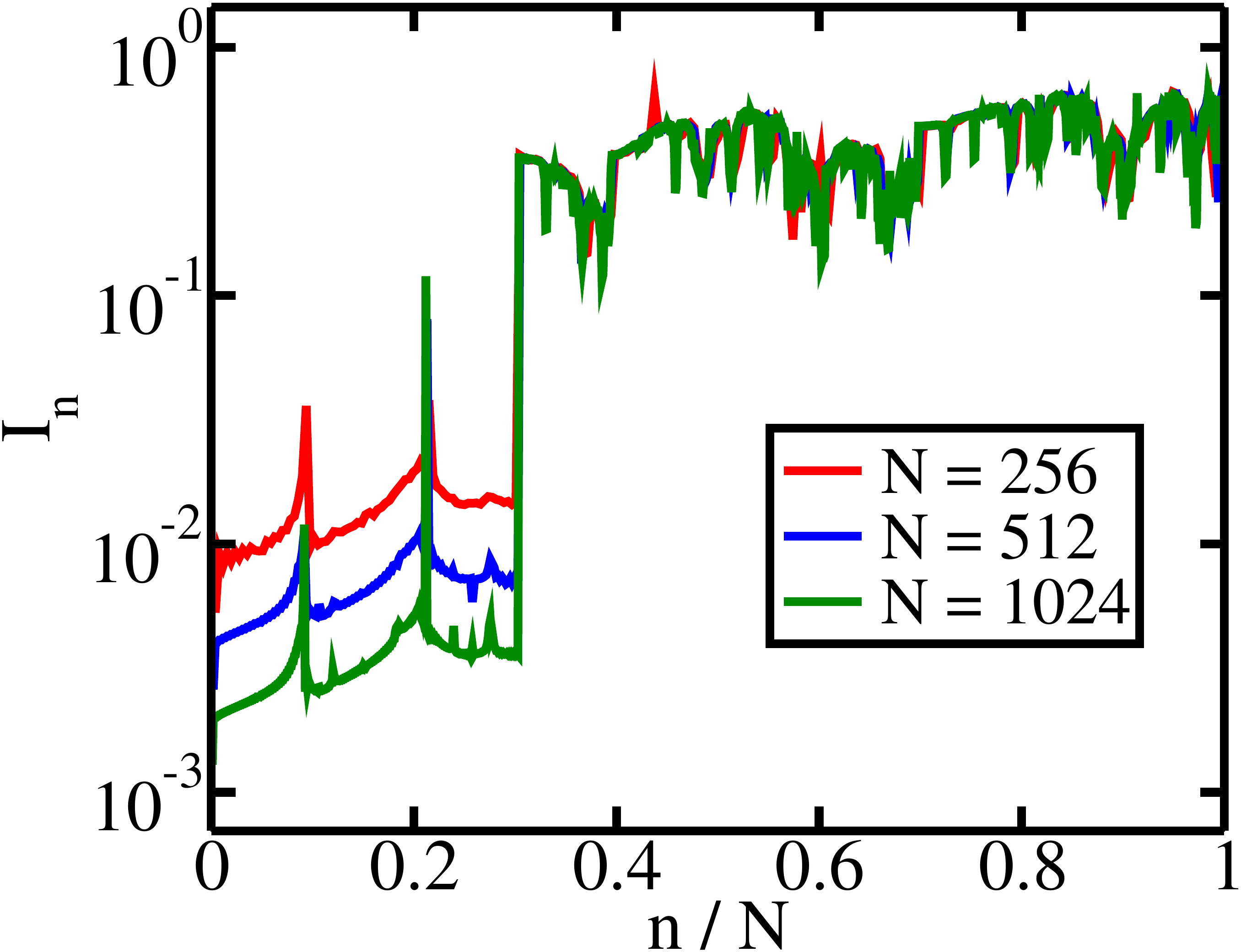}}{(h)}
	\stackunder{\includegraphics[width=5.2cm,height=4.1cm]{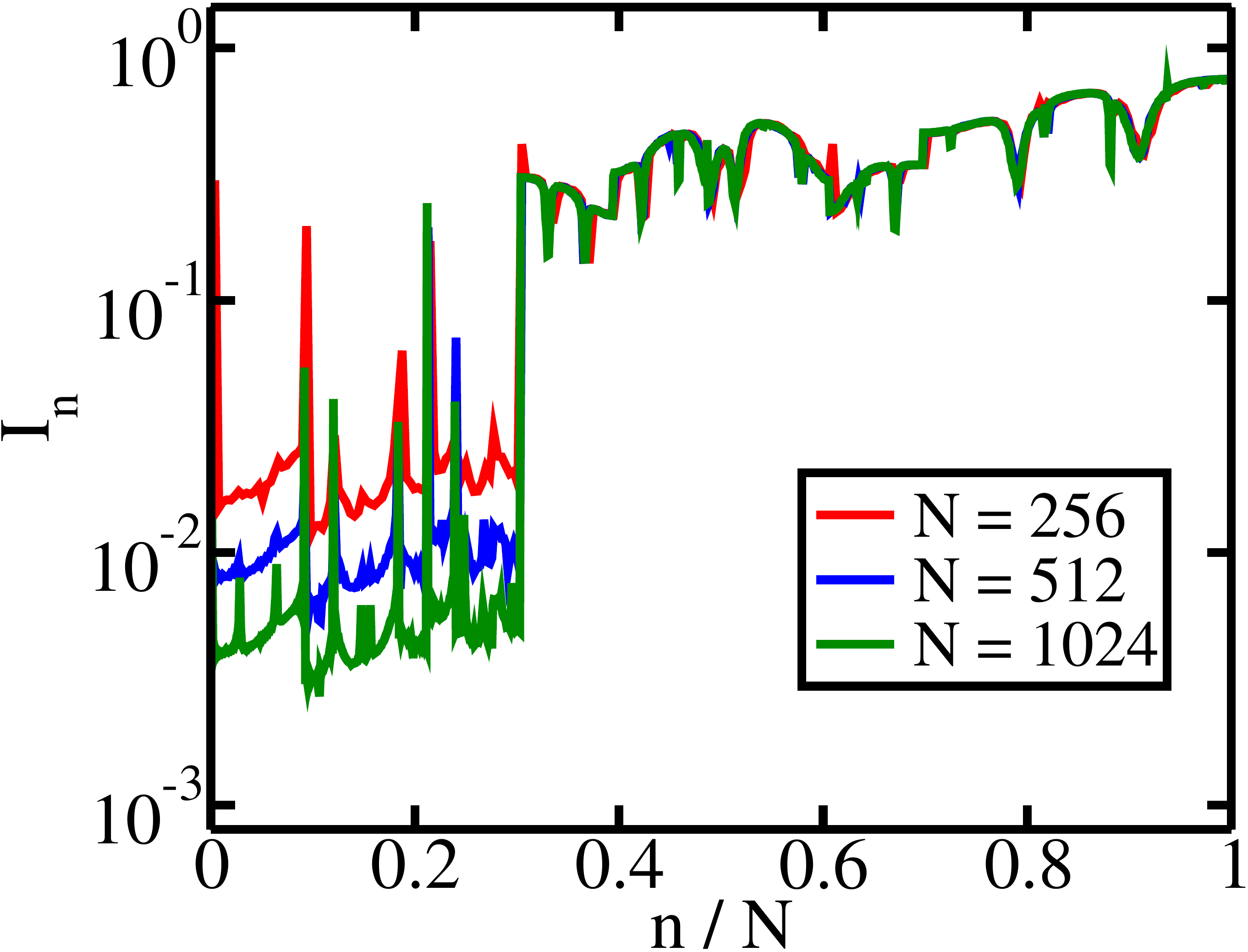}}{(k)}	
	\caption{(a-c) The inverse participation ratio $I_n$ of the single-particle eigenstates for $\alpha_g=(\sqrt{5}-1)/2$ with increasing system sizes $N=256,512, 1024$ for $\sigma=0.5, 1.5$ and $3.0$ respectively. (d-f) Similar plots for $\alpha_s=(\sqrt{2}-1)$ with increasing $N$ for $\sigma=0.5,1.5$ and $3.0$ respectively. (h-k) Similar plots for $\alpha_b=(\sqrt{13}-3)/2$ with increasing $N$ for $\sigma=0.5,1.5$ and $3.0$ respectively. For all the plots, $\lambda$ is kept fixed at $\lambda=2.2$. $n/N$ is the fractional eigenstate index.}
	\label{lrh_ipr}
\end{figure*} 

Fig.~\ref{D2_lrh}(b,c) for $\sigma=1.5$ and $3.0$ respectively show the appearance of blocks of localized states ($D_2\approx0$) with increasing $\lambda$. This implies that there exists a delocalized-to-localized (DL) edge, also well known as the mobility edge. Similar to DM edges these fixed DL-edge containing phases are also denoted as $P_q$ $(q=1,2,3,...)$ for $\eta=\alpha_g,\alpha_g^2,\alpha_g^3,...$ respectively.  $D_2$ of all the eigenstates for $\alpha_s$ and increasing $\lambda$ is shown in Fig.~\ref{D2_lrh}(d-f) for $\sigma=0.5,1.5,3.0$ respectively. For
$\alpha_s$ one obtains $P_1,P_2, P_3,...$ phases with $\eta=\alpha_s+\alpha_s^2,\alpha_s,\alpha_s^2 + \alpha_s^3,...$ and DM edges (for $\sigma=0.5$) and DL edges (for $\sigma=1.5,3.0$). Similarly from Fig.~\ref{D2_lrh}(g-k) for $\alpha_b$ and $\sigma=0.5,1.5,3.0$ respectively one obtains $P_1,P_2,P_3,...$ phases with $\eta=2\alpha_b+\alpha_b^2,\alpha_b+\alpha_b^2,\alpha_b,...$ and DM edges (for $\sigma=0.5$) and DL edges (for $\sigma=1.5,3.0$).

As evidence for multifractality we plot
$\langle D_f\rangle$ as a function of $f$ for the $P_2$ phase (with $\alpha_g^2$
fraction of delocalized states) for $\sigma=0.5$ (in
Fig.~\ref{Dq_lrh}(a)) and $\sigma=1.5$ (in Fig.~\ref{Dq_lrh}(b)) in
the LRH model with the `golden mean' $\alpha_g$. Here $\langle D_f\rangle$ denotes
$D_f$ averaged over $\alpha_g^2$ fraction of delocalized and
$(1-\alpha_g^2)$ fraction of non-delocalized eigenstates.  We chose a
Fibonacci system size $N=987$ to reduce the fluctuations due to high-$IPR$
eigenstates in the delocalized phase (see Appendix~\ref{app_aahstat}). In Fig.~\ref{Dq_lrh}(a) and
Fig.~\ref{Dq_lrh}(b) $\langle D_f\rangle$ averaged over $\alpha_g^2$ fraction of
eigenstates shows a similar small variation with $f$ with $\langle D_f\rangle$
being close to $1$, which implies these states are
delocalized. $\langle D_f\rangle$ averaged over $(1-\alpha_g^2)$ fraction of
eigenstates is a fraction and shows a non-trivial dependence on $f$
for $\sigma=0.5$ whereas $\langle D_f\rangle$ is close to $0$ and shows almost no
dependence on $f$ for $\sigma=1.5$. This indicates that these states
are multifractal for $\sigma=0.5$ and localized for
$\sigma=1.5$. Similar states can be found in the other $P_q$ phases
corresponding to $\alpha_s$ and $\alpha_b$.
\begin{figure*}
\centering
 			\stackunder{\includegraphics[width=0.63\columnwidth,height=4.3cm]{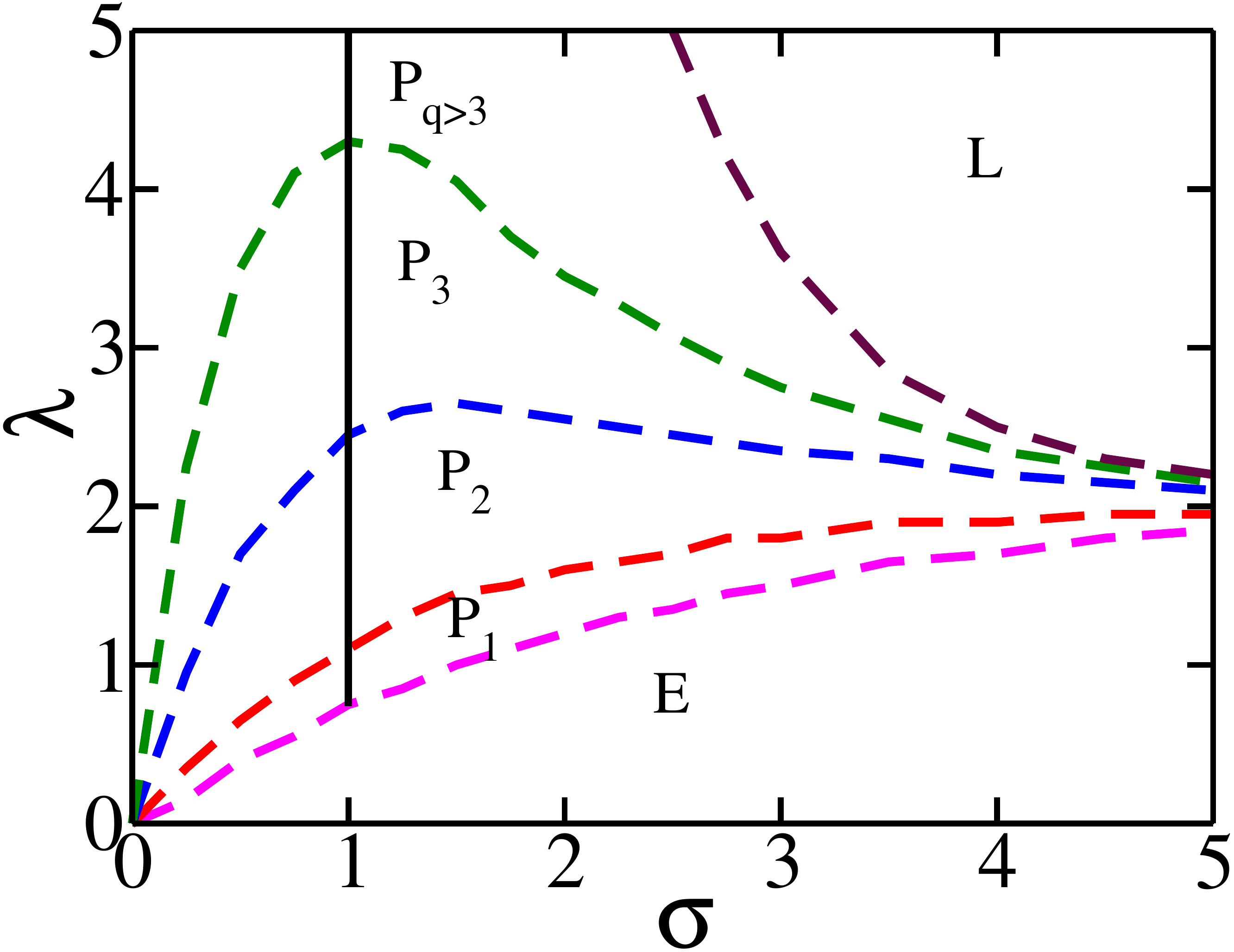}}{(a)}\hspace{0.25cm}
		  	\stackunder{\includegraphics[width=0.63\columnwidth,height=4.3cm]{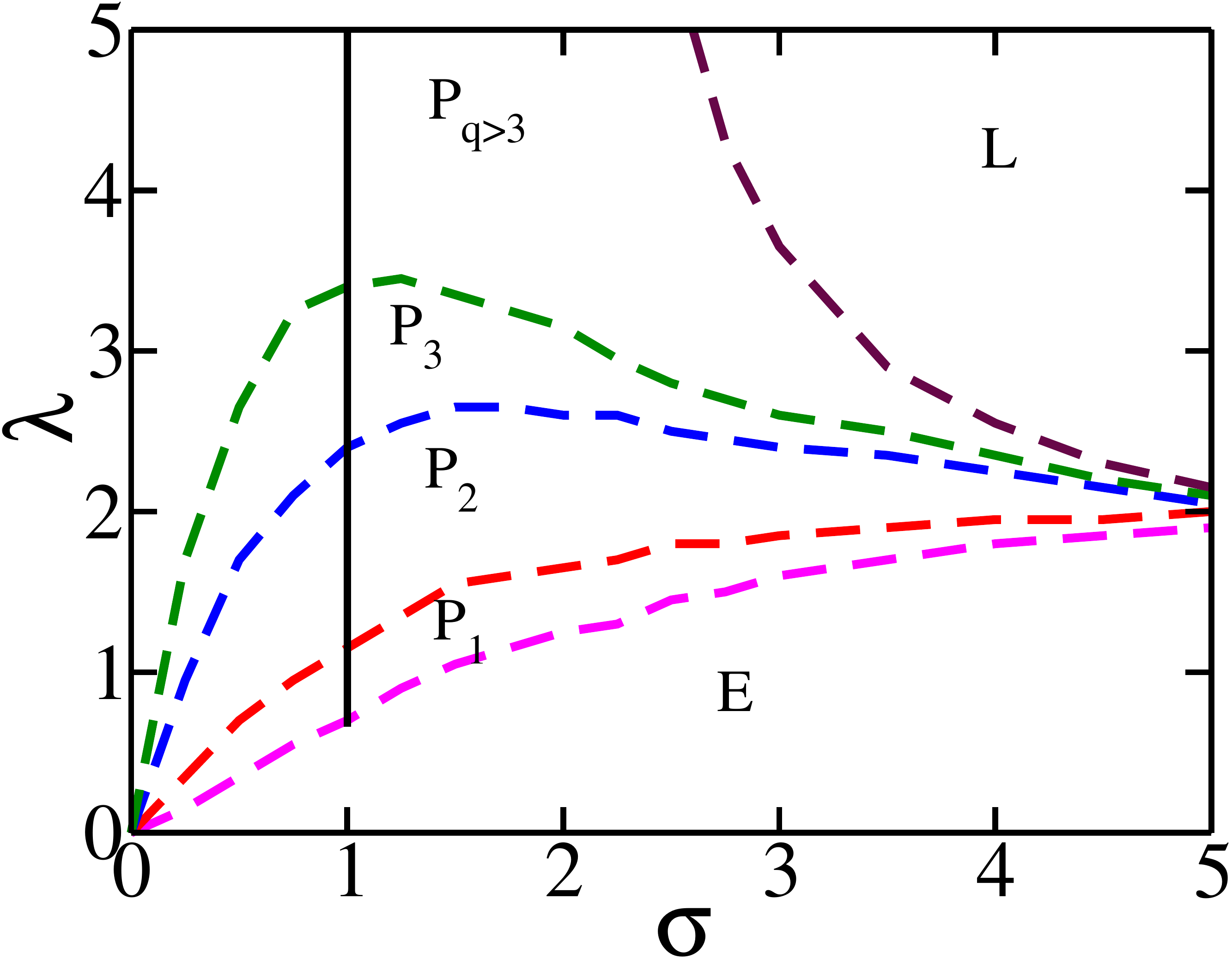}}{(b)}\hspace{0.25cm}
		  	\stackunder{\includegraphics[width=0.63\columnwidth,height=4.3cm]{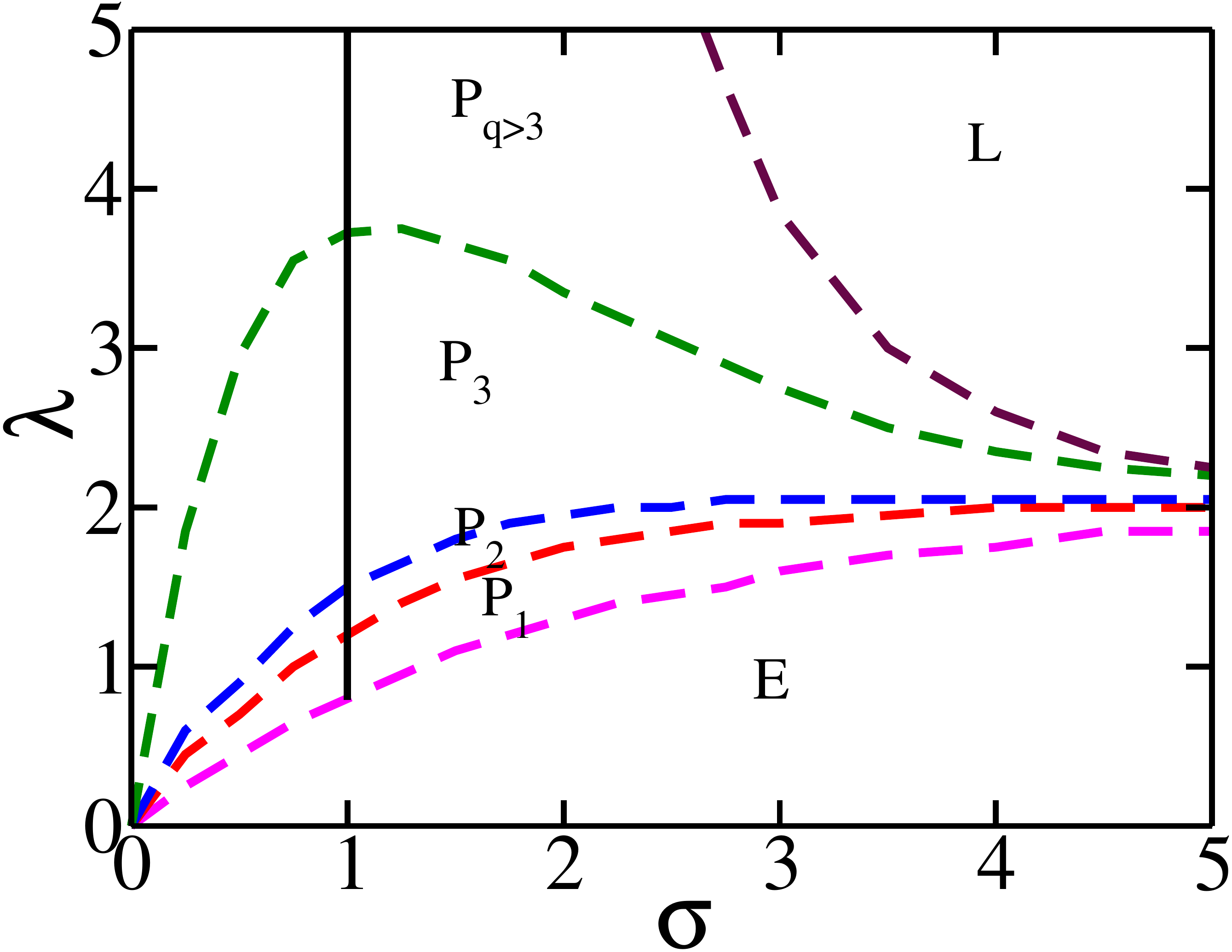}}{(c)}
	\caption{Phase diagram: in addition to
          extended ($E$) and localized ( $L$) phases with $\eta=1,0$
          respectively, presence of the mixed phases with fractional $\eta$: $P_1,P_2,P_3,...$ phases with (a)  $\eta=\alpha_g,\alpha_g^2,\alpha_g^3,...$; (b)
          $\eta=\alpha_s+\alpha_s^2,\alpha_s,\alpha_s^2+\alpha_s^3,...$; (c) $\eta=2\alpha_b^2+\alpha_b,\alpha_b^2 + \alpha_b,\alpha_b,...$. The
          vertical line separates out the DM edge for $\sigma<1$ from the DL
          edge for $\sigma>1$. Figures (a-c) are for $\alpha_g,\alpha_s,\alpha_b$ respectively.}
	\label{phase_lrh}
\end{figure*}

\subsection{Inverse participation ratio}
The inverse participation ratio ($IPR$) is a key quantity
for studying delocalization-localization transitions. It is defined as
\begin{eqnarray}
I_n = \sum_{i=1}^{N} |\psi_n(i)|^4,
\end{eqnarray}
where the $n\textsuperscript{th}$ normalized single particle
eigenstate $\ket{\psi_n}=\sum_{i=1}^{N}\psi_n(i) \ket{i}$ is written
in terms of the Wannier basis $\ket{i}$, representing the state of a
single particle localized at the site $i$ of the lattice.  For a
delocalized eigenstate $I_n\propto N^{-1}$ whereas for a localized
eigenstate $I_n\propto N^0$. For a critical state $I_n$ shows
intermediate behavior. 
Here we calculate $IPR$ of the eigenstates for the LRH model with finite $\sigma$.
To get a hint about the phases in the model, here we choose a fixed
$\lambda=2.2$ (which corresponds to the localized phase in the $\sigma\rightarrow\infty$ limit)
and different values of $\sigma=0.5,1.5,3.0$ for quasiperiodicity parameters $\alpha_g, \alpha_s$ and $\alpha_b$. The $IPR$ of all the single particle eigenstates for
$\alpha_g$ are shown in Fig.~\ref{lrh_ipr}(a),(b) and (c) for
$\sigma=0.5, 1.5$ and $3.0$ respectively.
Fig.~\ref{lrh_ipr}(a) shows
that the eigenstates are delocalized $(I_n\propto N^{-1})$ as long as the fractional index
$n/N<\alpha_g^3$. The $IPR$ of the remaining eigenstates for
$n/N>\alpha_g^3$ shows an intermediate dependence on $N$
i.e. $N^{-1}<I_n<N^0$. It turns out that these eigenstates are
multifractal~\cite{deng2019one}(see Fig.~\ref{Dq_lrh}(a)). Hence a DM edge exists at $n/N=\alpha_g^3$ for
$\sigma=0.5$ and $\lambda=2.2$. As shown in Fig.~\ref{lrh_ipr}(b) and
Fig.~\ref{lrh_ipr}(c) the eigenstates are delocalized for
$n/N<\alpha_g^2$ whereas the eigenstates are localized ($I_n\propto
N^0$) for $n/N>\alpha_g^2$ for the same $\lambda$ and $\sigma=1.5$ and $3.0$
respectively. This implies that there exists a DL
(mobility) edge for $\lambda=2.2$ and $\sigma=1.5, 3.0$. Also we
notice that the fraction of the delocalized eigenstates can change
with $\sigma$ for a fixed $\lambda$. However, the occasional presence of the high-$IPR$ states as discussed for the AAH model (see Appendix~\ref{app_aahstat}), especially in the
delocalized regime are also visible for the LRH model, since values of $N$ are
chosen to be non-Fibonacci numbers in all the plots of
Fig.~\ref{lrh_ipr} .

We also show the results obtained from the LRH model for the silver
and bronze means. Plots obtained using $\alpha_s$ and $\lambda=2.2$
are shown in Fig~\ref{lrh_ipr}(d-f) for $\sigma=0.5,1.5$ and $3.0$
respectively. These figures indicate that there is a DM edge at
$n/N\approx\alpha_s^2+\alpha_s^3$ for $\sigma=0.5$ whereas there is a
DL edge at $n/N\approx\alpha_s$ for $\sigma=1.5$ and $3.0$.
Fig.~\ref{lrh_ipr}(g-k) are obtained using fixed $\alpha_b$,
$\lambda=2.2$ for $\sigma=0.5,1.5$ and $3.0$ respectively. It can be
seen from Fig.~\ref{lrh_ipr}(g-k) that there is a DM edge at
$n/N\approx\alpha_b$ for $\sigma=0.5$ whereas a DL edge exists at
$n/N\approx\alpha_b$ for $\sigma=1.5$ and $3.0$. We see that in every
plot of Fig.~\ref{lrh_ipr} the fraction of delocalized eigenstates can
always be expressed as a function of the parameter $\alpha$.  However,
the $IPR$ fluctuations due to the presence of the high-$IPR$ states in the delocalized regime continue to persist
in these cases also, although they tend to vanish if $N$ is a Fibonacci
number as can be seen in the AAH model (see Appendix~\ref{app_aahstat}). It is noticeable that the $IPR$ fluctuations increase in the
delocalized regime with $\sigma$.

\subsection{Phase diagram}
After an extensive analysis, we find that in a particular $P_q$ phase, the same blocks of multifractal states become
localized as one crosses $\sigma=1$ whereas the corresponding $\eta$ remains the same.  We chart out the single-particle
phase diagram for the parameter $\alpha_g$ in
Fig~\ref{phase_lrh}(a), which is also obtained in
Ref.~\onlinecite{deng2019one} for a Fibonacci
$N$. Fig~\ref{phase_lrh}(a-c) contain special states with high-$IPR$ eigenstates, similar to the AAH model (see Appendix~\ref{app_aahstat}), even in the delocalized regimes. It is to be noted that as
$\sigma$ increases the extent of the mixed phases shrinks as the LRH model
approaches the AAH limit.  The phase diagrams for $\alpha_s$ and
$\alpha_b$ are shown in Fig.~\ref{phase_lrh}(b) and
Fig.~\ref{phase_lrh}(c) respectively. The $P_q$ phases
(corresponding to $\alpha_s$ and $\alpha_b$) in these cases as well,
like with $\alpha_g$, contain DM edges for $\sigma<1$ and DL edges for
$\sigma>1$. The changes in $P_q$ phases at $\sigma=1$ are denoted by
the vertical lines in all the phase diagrams.
\begin{figure}
\centering
	\includegraphics[width=3.0cm,height=8.0cm]{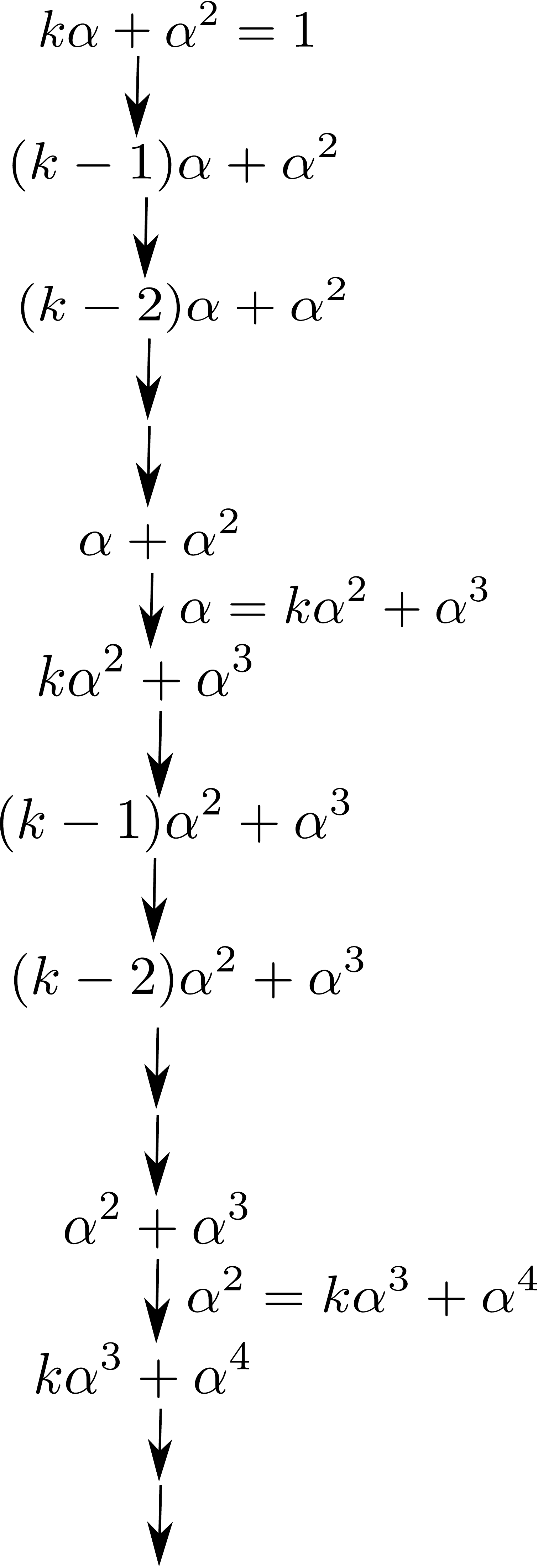}
	\caption{Depicts how the fraction of delocalized eigenstates
          ($\eta$) decreases in a manner that uses the rule defined in
          Eq.~\ref{fraction}. One can express the fraction of the
          delocalized states as a sum of two bits $k\alpha^{z+1}$ and
          $\alpha^{z+2}$, out of which the bigger bit loses weight at
          every step until it reaches $\alpha^{z+1}$, where it
          disintegrates according to the rule defined in
          Eq.~\ref{fraction} and then the bigger bit loses weight at
          each step. For a specific value of $\alpha$, at every step
          of the sequence one obtains a $P_q$ phase.}
	\label{seq}
\end{figure}

\section{Fraction of delocalized states}\label{lrh_sec3}
After a careful observation of the phase diagrams, one may propose a sequence which
dictates the values of $\eta$ in $P_q$
phases corresponding to different quasiperiodicity parameters $\alpha$,
which belong to the metallic mean family described in Eq.~\ref{fibo}. 
For any $\sigma>0$ without disorder $(\lambda=0)$, $\eta=k\alpha +
\alpha^2=1$ where $k=1,2,3$ correspond to $\alpha_g,\alpha_s,\alpha_b$
respectively and $z=0$ in Eq.~\ref{fraction}.  As the quasiperiodic
disorder is turned on $(\lambda\neq0)$, $\eta$ starts decreasing in a
sequence according to Eq.~\ref{fraction} for the metallic mean
family, which is depicted in Fig.~\ref{seq} . Eq.~\ref{fraction}
implies that one can always express $(\alpha)^z$ as a sum of two bits
$k(\alpha)^{z+1}$ and $(\alpha)^{z+2}$. In the LRH model the bigger
bit loses weight at every step becoming $(k-1)\alpha^{z+1},
(k-2)\alpha^{z+1},...$ until it reaches $\alpha^{z+1}$, where it
disintegrates again according to the rule defined in
Eq.~\ref{fraction} and the new bigger bit starts losing weight at each
step. This is a continuous process as depicted by the sequence in
Fig.~\ref{seq} . For a specific choice of $\alpha$, one obtains a
$P_q$ phase at each step of the sequence. The top of the sequence
corresponds to the fully delocalized $(\eta=1)$ phase. One obtains
$P_1,P_2,..$ phases as one goes down following the sequence. The $P_q$
phases possess DM (DL) edges if $\sigma<1$ ($\sigma>1$).

We show a schematic of the phase diagram in Fig.~\ref{schematic_lrh},
where the colored regions are labelled by $\eta$ in different
phases. Choosing $k=2$ in the sequence depicted in Fig.~\ref{seq}
leads to the phases labelled by $\eta$ in Fig.~\ref{schematic_lrh}
. These phases are as follows $\rightarrow$ red:
$\eta=2\alpha+\alpha^2=1$ (delocalized); green: $\eta=\alpha+\alpha^2$
($P_1$); orange: $\eta=2\alpha^2+\alpha^3$ ($P_2$); purple:
$\eta=\alpha^2+\alpha^3, 2\alpha^3+\alpha^4,...$ ($P_3, P_4...$
respectively) collectively, which appear as one proceeds further
according to the sequence. For large values of $\sigma$ and $\lambda$
the localized phase appears when $\eta=0$, shown in blue.

\section{Entanglement entropy}\label{lrh_sec4}
Here we consider noninteracting spinless
fermions in the LRH model to calculate the entanglement entropy of the
fermionic ground states in different phases obtained in the previous
section.  The entanglement entropy in the ground state of such free
fermionic systems is given
by~\cite{peschel2003calculation,peschel2009,peschel2012special}
\begin{equation}
S_A=-\sum\limits_{m=1}^{L} [\zeta_m \log \zeta_m + (1-\zeta_m) \log (1-\zeta_m)],
\end{equation}
where $\zeta_m$'s are the eigenvalues of the correlation matrix $C^A$,
where $C^A_{ij}=\left\langle c_{i}^{\dagger}c_{j} \right\rangle$ with
$i,j\in$ subsystem $A$ of $L$ sites. For free fermions in $d$
dimensions, typically $S_A\propto L^{d-1}\ln L$ in metallic
phases~\cite{swingle}, while it goes as $S_A\propto L^{d-1}$ in
adherence to the `area-law' in the localized phases in the presence of
disorder.
\begin{figure}
\centering
	\stackunder{\includegraphics[width=0.45\columnwidth,height=3.1cm]{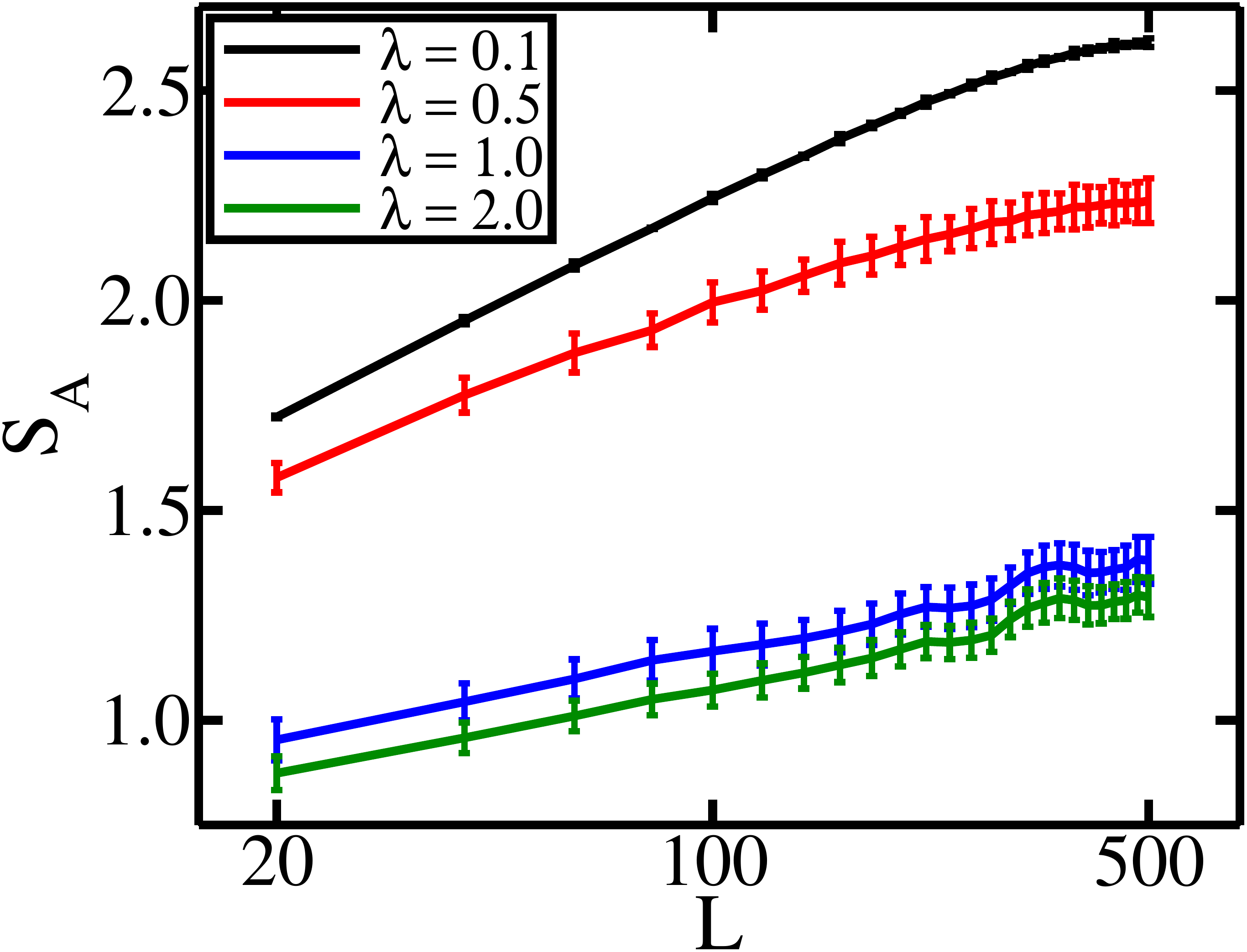}}{(a)}
	\stackunder{\includegraphics[width=0.45\columnwidth,height=3.1cm]{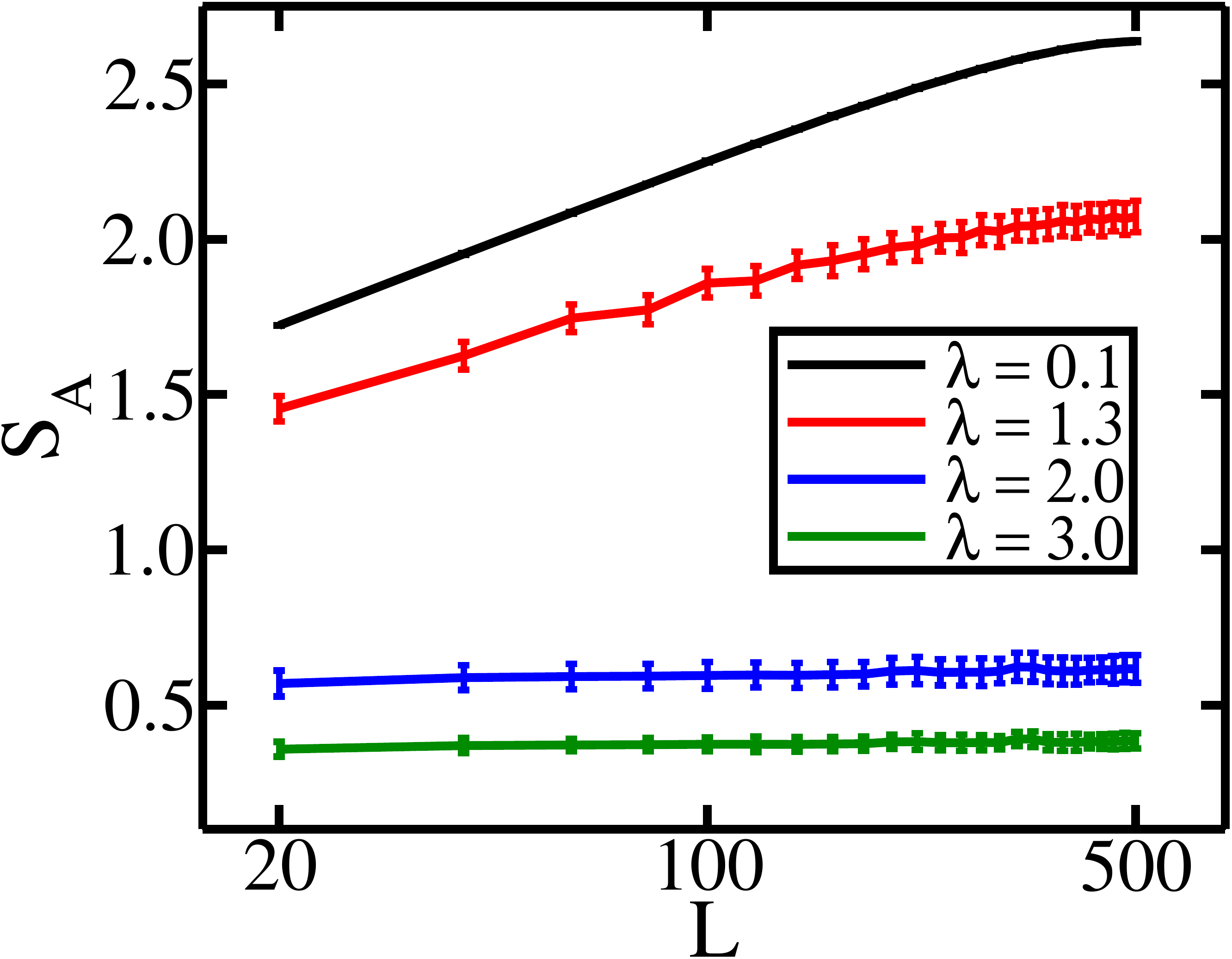}}{(b)}
	\stackunder{\includegraphics[width=0.45\columnwidth,height=3.1cm]{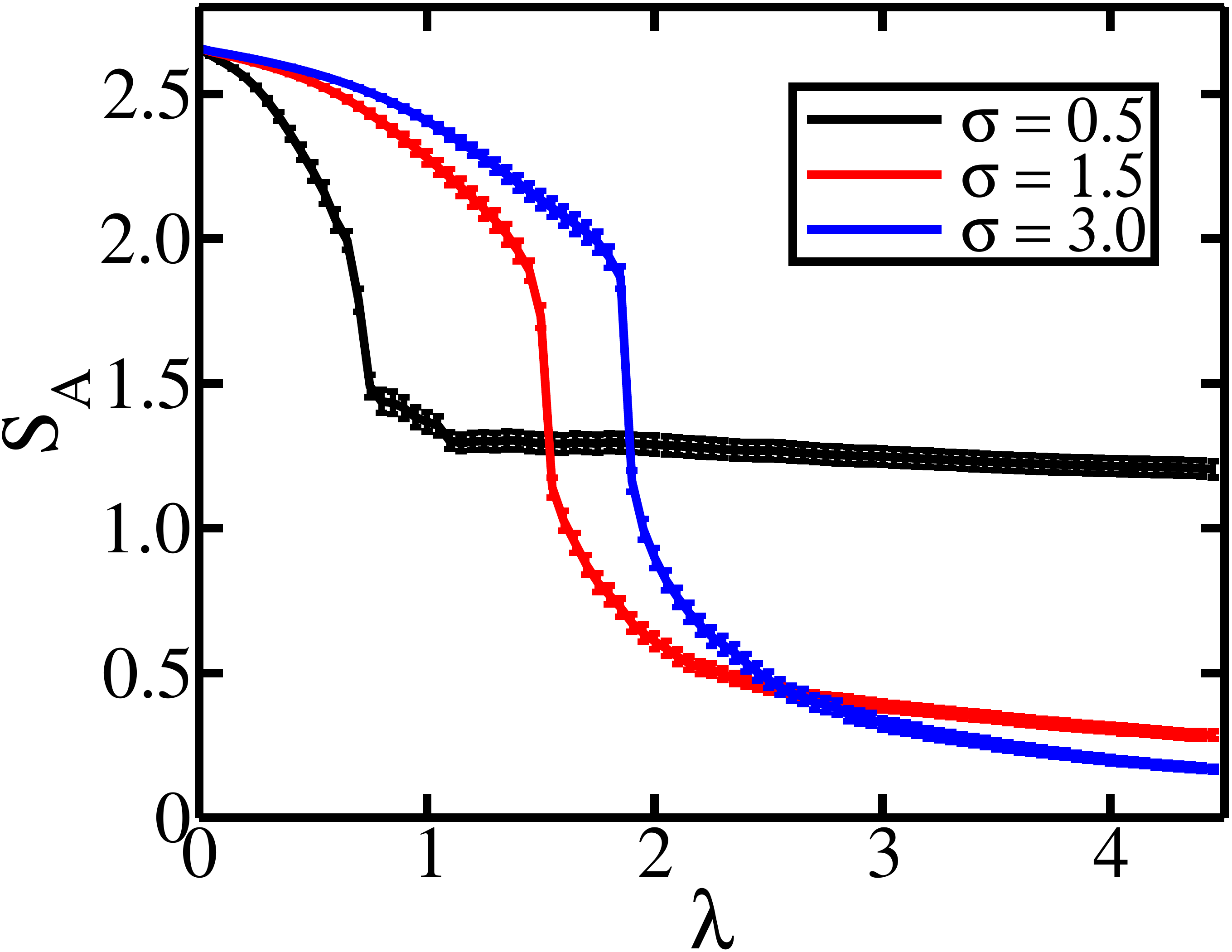}}{(c)}
	\stackunder{\includegraphics[width=0.45\columnwidth,height=3.1cm]{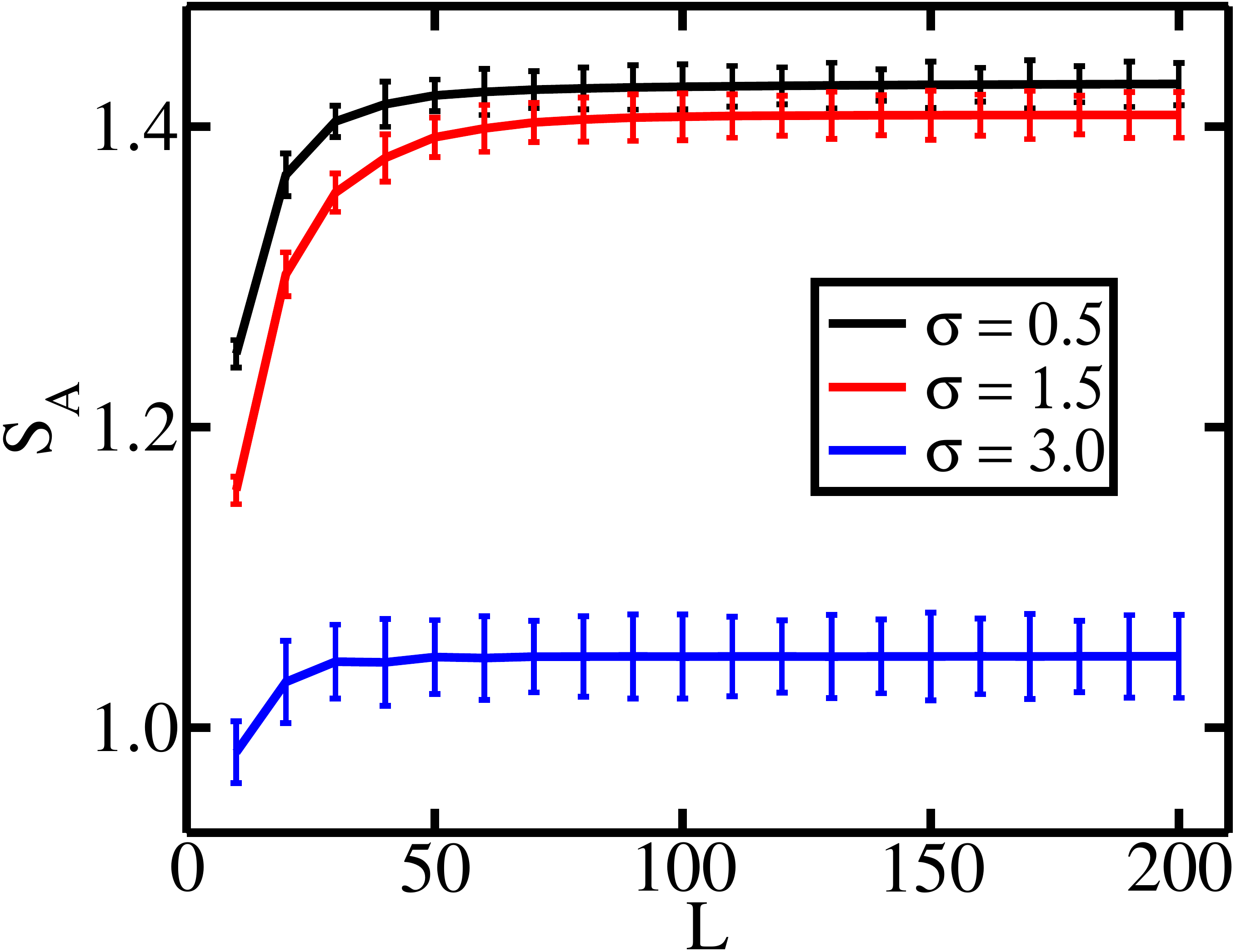}}{(d)}
	\caption{(a-b) The subsystem size $L$ dependence of entanglement entropy $S_A$ with increasing values of $\lambda$ for fermions at half-filling and for $\sigma=0.5$ and $1.5$. (c) $S_A$ as function of $\lambda$ for $\sigma=0.5,1.5$ and $3.0$ respectively for fermions at half-filling with $L=N/2$. For all the plots in figures (a-c) $N=1024$. (d) Entanglement entropy $S_A$ as a function of subsystem size $L$ for increasing $\sigma$ and fixed $\lambda=2.2$. For all the plots $N=512$  for $\alpha_g$ and special filling $\nu=\alpha_g^4$.}
	\label{eevsL_lrh}
\end{figure}

To produce smoother plots, we employ an average of $S_A$ over the
$100$ realizations of $\theta_p$ uniformly choosing from $[0,2\pi]$ in
all the plots here. We stick to filling fraction $\nu=0.5$ of fermions
unless otherwise mentioned and $\alpha_g=(\sqrt{5}-1)/2$.  The $S_A$
vs $L$ plots are shown in Fig.~\ref{eevsL_lrh}(a) and
Fig.~\ref{eevsL_lrh}(b) at half-filling with increasing values of
$\lambda$ for $\sigma=0.5$ and $1.5$ respectively. 
A generic scaling form $S_A=K\ln L + K_0$ is assumed for this purpose. From
Fig.~\ref{eevsL_lrh}(a) for $\sigma=0.5$, when $\lambda=0.1$ and $0.5$
(delocalized and $P_1$ phases), the Fermi level is delocalized and
hence $S_A\propto\ln L$ with $K\approx0.32$ and $0.25$ respectively. In the same figure, when $\lambda=1.0$ and
$2.0$ ($P_2$ and $P_3$ phases) the Fermi level is multifractal,
$S_A\propto \ln L$ but the magnitude of $S_A$ is drastically low with $K\approx0.13$ for both the cases.
The logarithmic scaling behavior indicates that the multifractal states are essentialling extended states but with nonergodicity as the prefactor differs from the ergodic extended/delocalized states. 
In Fig.~\ref{eevsL_lrh}(b) for $\sigma=1.5$, when $\lambda=0.1$ and $1.3$
(delocalized and $P_1$ phases), the Fermi level is delocalized and
$S_A\propto\ln L$ with $K\approx0.33$ and $0.24$ respectively. However, when $\lambda=2.0$ and $3.0$ ($P_2$ and
$P_3$ phases) the Fermi level is localized, the magnitude of $S_A$ is
much lower, and it abides by the area-law $(K\approx0)$. Transitions of Fermi level
at half-filling are shown in Fig.~\ref{eevsL_lrh}(c) for
$\sigma=0.5,1.5$ and $3.0$ respectively. For $\sigma=0.5$, the Fermi
level undergoes a DM transition at $\lambda=0.75$. For $\sigma=1.5$
and $3.0$ the Fermi level undergoes DL transitions at $\lambda=1.5$
and $\lambda=1.85$ respectively, as also evident from
Fig.~\ref{D2_lrh}(a-c).

We have also checked that the qualitative behavior of $S_A$ vs $L$ plots in
the half-filled free fermionic ground state barely changes in the
phase diagram for $\alpha_s$ and $\alpha_b$. However, similar to the
AAH model~\cite{roy2019study} (see Appendix~\ref{app_aahstat}), the LRH model too shows `area-law' behavior for special fillings $\nu$ even in the delocalized
regime.  An example of this is shown in Fig.~\ref{eevsL_lrh}(d) for
$\lambda=2.2$ and $\sigma=0.5,1.5,3.0$ and special filling
$\nu=\alpha_g^4$. In all these plots $S_A$ abides by the
`area-law'. However, the magnitude of $S_A$ is significantly smaller
for $\sigma=3.0$. We point out that while the single particle results
depend on whether the system size is a Fibonacci number, the
many-particle measures do not show such a dependence on the system
size (see Appendix~\ref{app_aahstat}) for
details).
\begin{figure}
\centering
	\stackunder{\includegraphics[width=0.49\columnwidth,height=3.7cm]{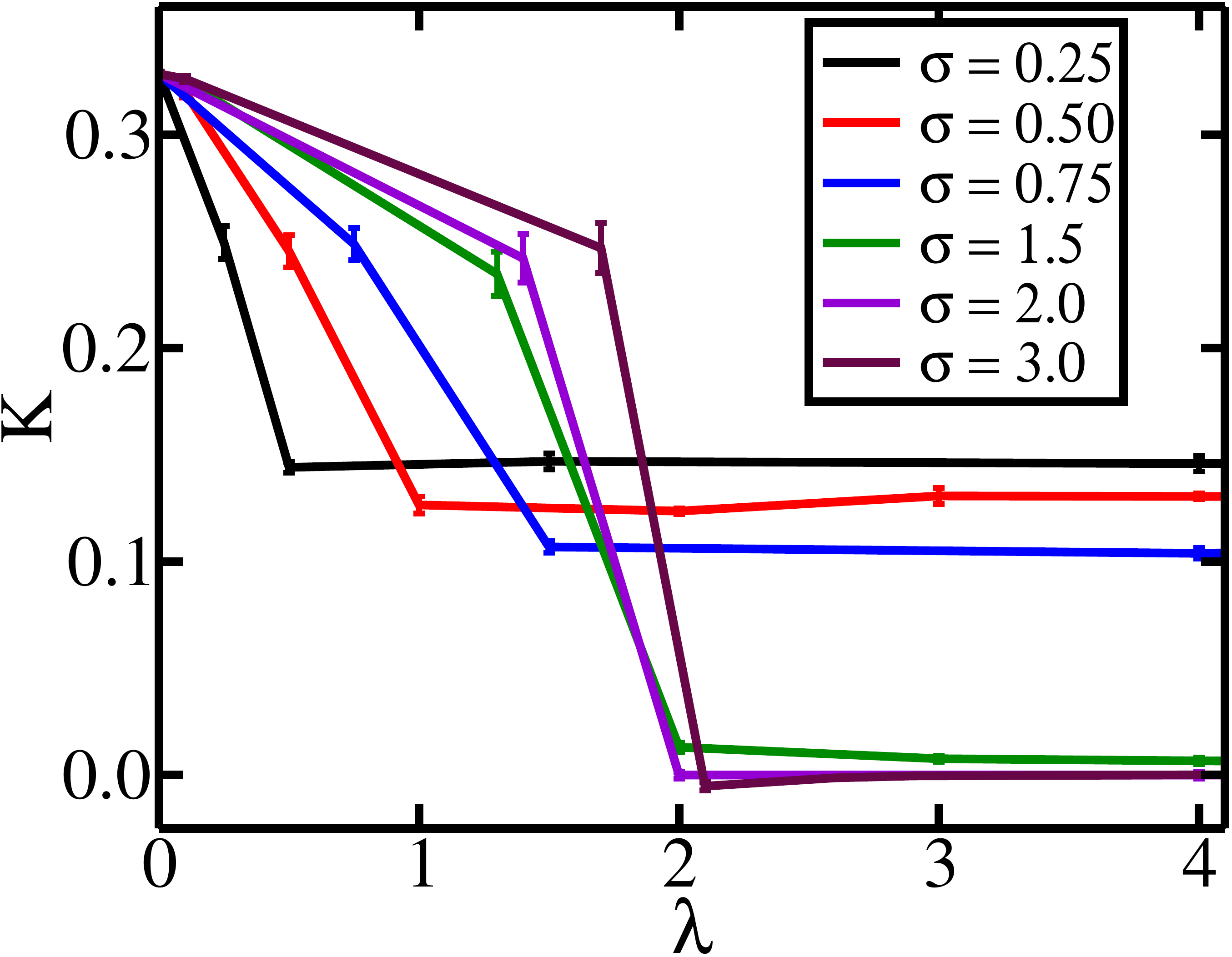}}{(a)}
	\stackunder{\includegraphics[width=0.49\columnwidth,height=3.7cm]{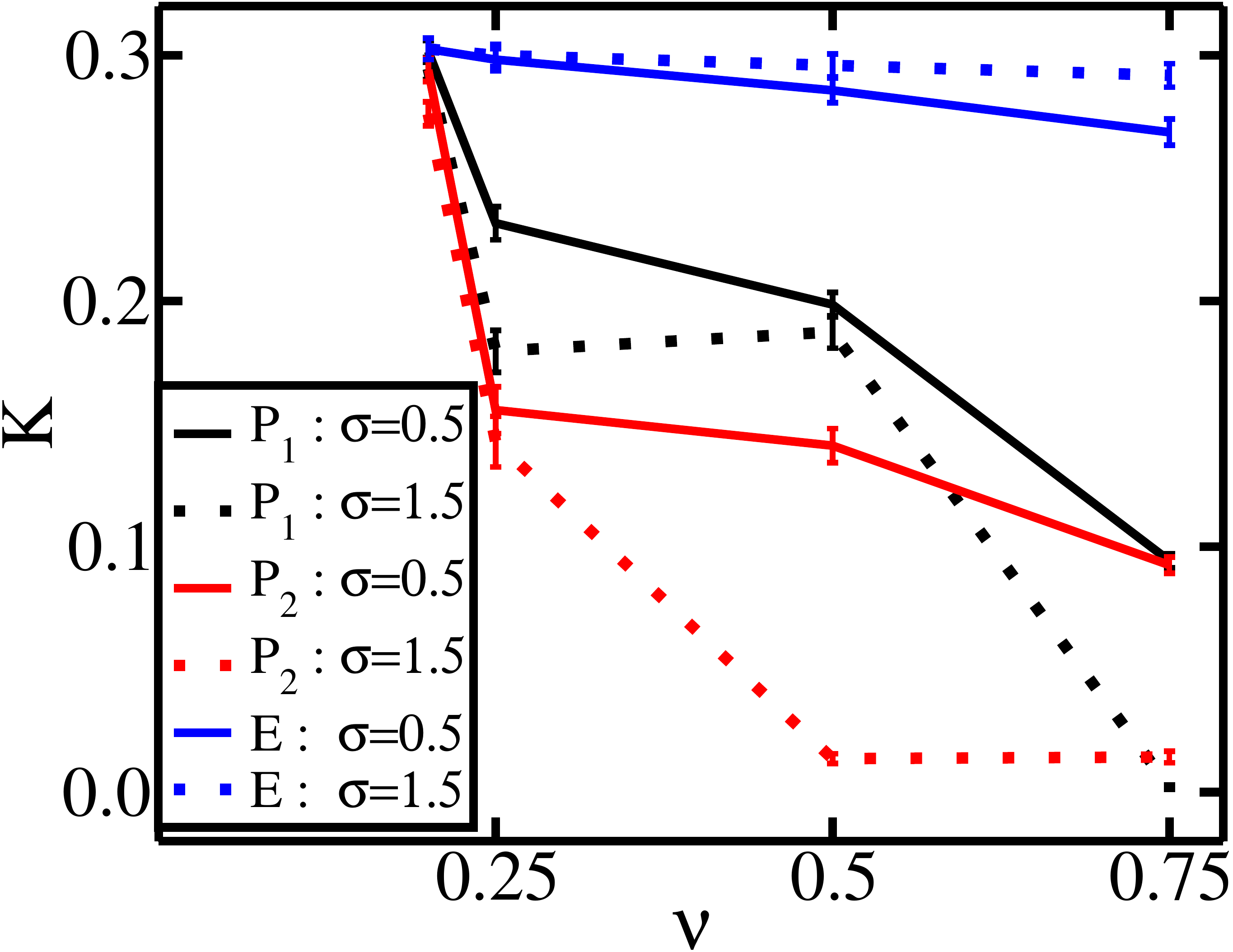}}{(b)}
	\caption{(a) The prefactor $K$ of the logarithmic term as a function of $\lambda$ for fermions at half-filling and for increasing values of $\sigma$.
	(b) $K$ as a function of (non-special) filling fraction $\nu$ in the mixed phases $P_1$, $P_2$ and extended/delocalized ($E$) phase for $\sigma=0.5$ and $1.5$.
	For all the plots $N=1024$ for $\alpha_g$.}
	\label{cofactor_lrh}
\end{figure}

Next we further analyze the prefactor $K$ of the log term from the subsystem size dependence of $S_A$ for $\alpha_g$. In Fig.~\ref{cofactor_lrh}(a), we show $K$ as a function of $\lambda$ for increasing values of $\sigma$ for fermionic ground state at half-filling. In the delocalized phase of a clean system for $\sigma>0$, $K\approx0.33$. With turning on of $\lambda$ the Fermi level at half filling becomes multifractal and localized at the $P_2$ phase for $\sigma<1$ and $\sigma>1$ respectively. In the $P_1$ phase, although being a mixed phase, the Fermi level and all other states below it remain delocalized at half-filling with the values of $K$ lying around $0.25$. This shows that the nature of the delocalized states change with $\lambda$. In $P_2$ phase the values of $K$ for multifractal Fermi level (for $\sigma<1$) decrease with $\sigma$ showing the change in multifractality of the Fermi level with $\sigma$. $K\approx0$ for $\sigma>1$ as the Fermi level gets localized at half-filling. In the $\sigma=\infty$ limit $K\approx0.33, 0.26$ and $0$ in the delocalized phase, at the critical point and in the localized phase of the AAH model~\cite{roosz2020entanglement,prb}. In Fig.~\ref{cofactor_lrh}(b) we show $K$ as a function of the non-special values of filling fraction $\nu$. The plots show that in the extended/delocalized $(E)$ phase $K$ depends very little on $\nu$ whereas in the mixed phases $(P_1, P_2)$ $K$ may depend significantly on $\nu$ as shown in the figure for $\sigma=0.5$ and $1.5$.       

\section{Conclusions}\label{lrh_sec5}
 We uncover an intricate pattern of the localization
structure of the AAH potential in the presence of long-range hoppings
when the quasiperiodicity parameter is a member of the `metallic mean
family'. In addition to the fully delocalized and localized phases we
obtain a co-existence of multifractal (localized) eigenstates with
delocalized eigenstates for $\sigma<1$ ($\sigma>1$). The fraction of
delocalized eigenstates in these phases can be obtained from a general
sequence which is a manifestation of a mathematical property of the
`metallic mean family'. The entanglement entropy of a noninteracting
fermionic ground state respects the area-law if the Fermi level
belongs in the localized regime while logarithmically violating it if
the Fermi-level belongs in the delocalized or multifractal regimes,
although the magnitude in the multifractal regime is significantly
lower than in the delocalized one. 
A study of the prefactor of the logarithmically violationg term in the subsystem size scaling of entanglement entropy shows interesting behavior in different phases.
The entanglement entropy surprisingly
follows the area-law for certain special filling fractions even in the
delocalized regime. These special filling fractions are related to the
metallic means.  In this work, we make an attempt to show how the
inherent mathematical structure in the metallic means manifests
itself in the single particle and many particle properties of a class
of quasiperiodic models. Studies of this kind are very rare in the
literature~\cite{thiem2009wave,thiem2011generalized}. Hopefully our
work will help motivate further research in this direction.
     
\section*{Acknowledgments} 
N. R is grateful to the University Grants
Commission (UGC), India for providing a PhD fellowship. A.S
acknowledges financial support from SERB via the grant (File Number:
CRG/2019/003447), and from DST via the DST-INSPIRE Faculty Award
[DST/INSPIRE/04/2014/002461].

\bibliography{refs} 

\appendix

\begin{center}
{\bf APPENDIX}
\end{center} 

In this section we
discuss the results involving the inverse participation ratio $(IPR)$,
fractal dimension and entanglement entropy of the AAH model with
nearest-neighbor hopping ($\sigma\rightarrow\infty$ limit of the LRH
model) and quasiperiodic potential. 

\section{IPR, fractal dimension and entanglement entropy in the AAH model}\label{app_aahstat}
The AAH model has a self-dual point at $\lambda=2$, where the
Hamiltonian in position space maps to itself in momentum space.  As a
consequence all the single-particle eigenstates are delocalized for
$\lambda<2$ and localized for $\lambda>2$~\cite{mathieu,aubry}. But
earlier studies~\cite{ASinha2018,roy2019study} of the same model based
on the golden mean as the quasiperiodicity parameter have shown the
existence of energy-dependent localization properties. Here we extend
the study to the case of metallic means.  We discuss the results for
various quantities ahead. \\

\begin{figure}
	\centering
	        \stackunder{\includegraphics[width=4.2cm,height=3.65cm]{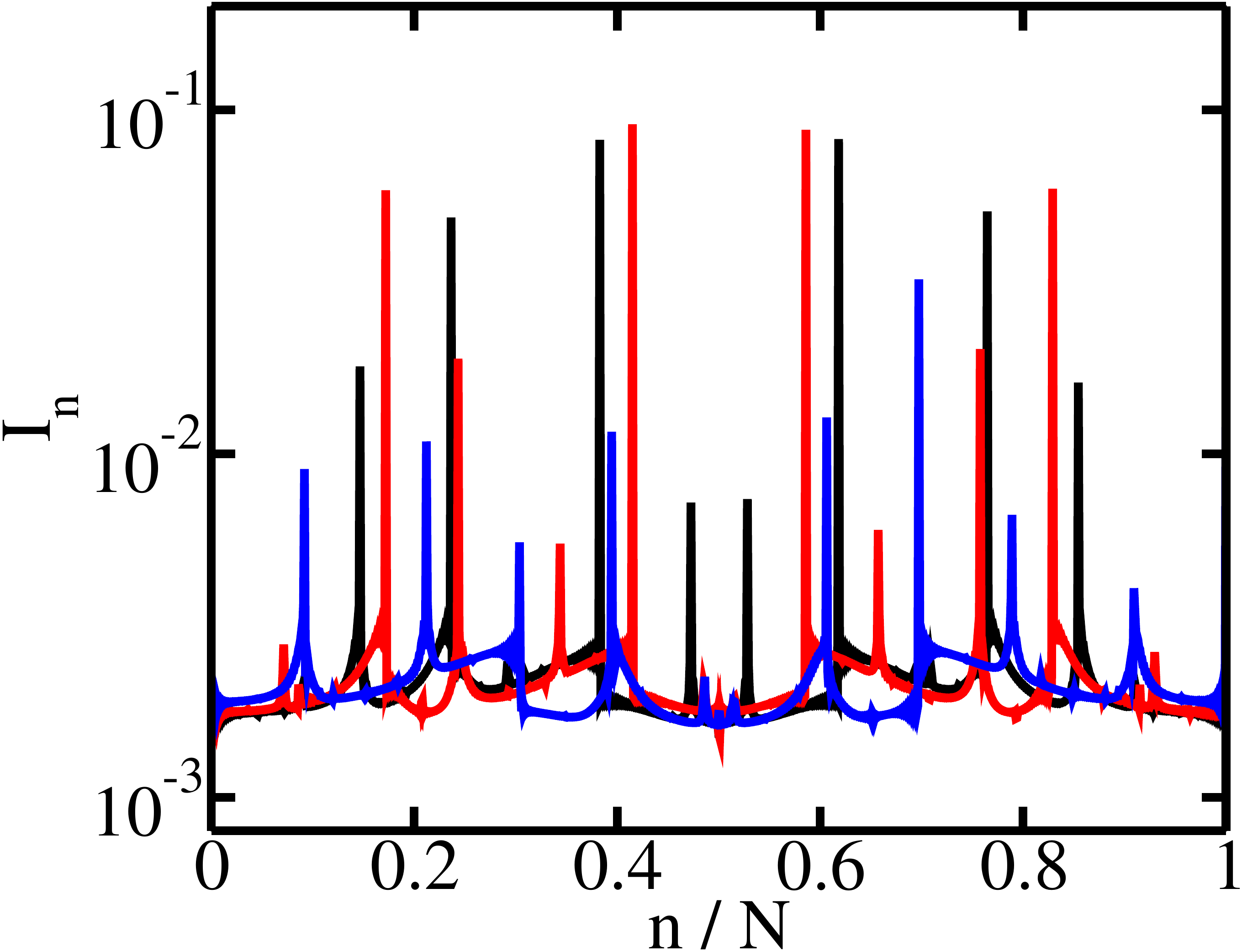}}{(a)}\hspace{0.15cm}
		  	\stackunder{\includegraphics[width=4.2cm,height=3.65cm]{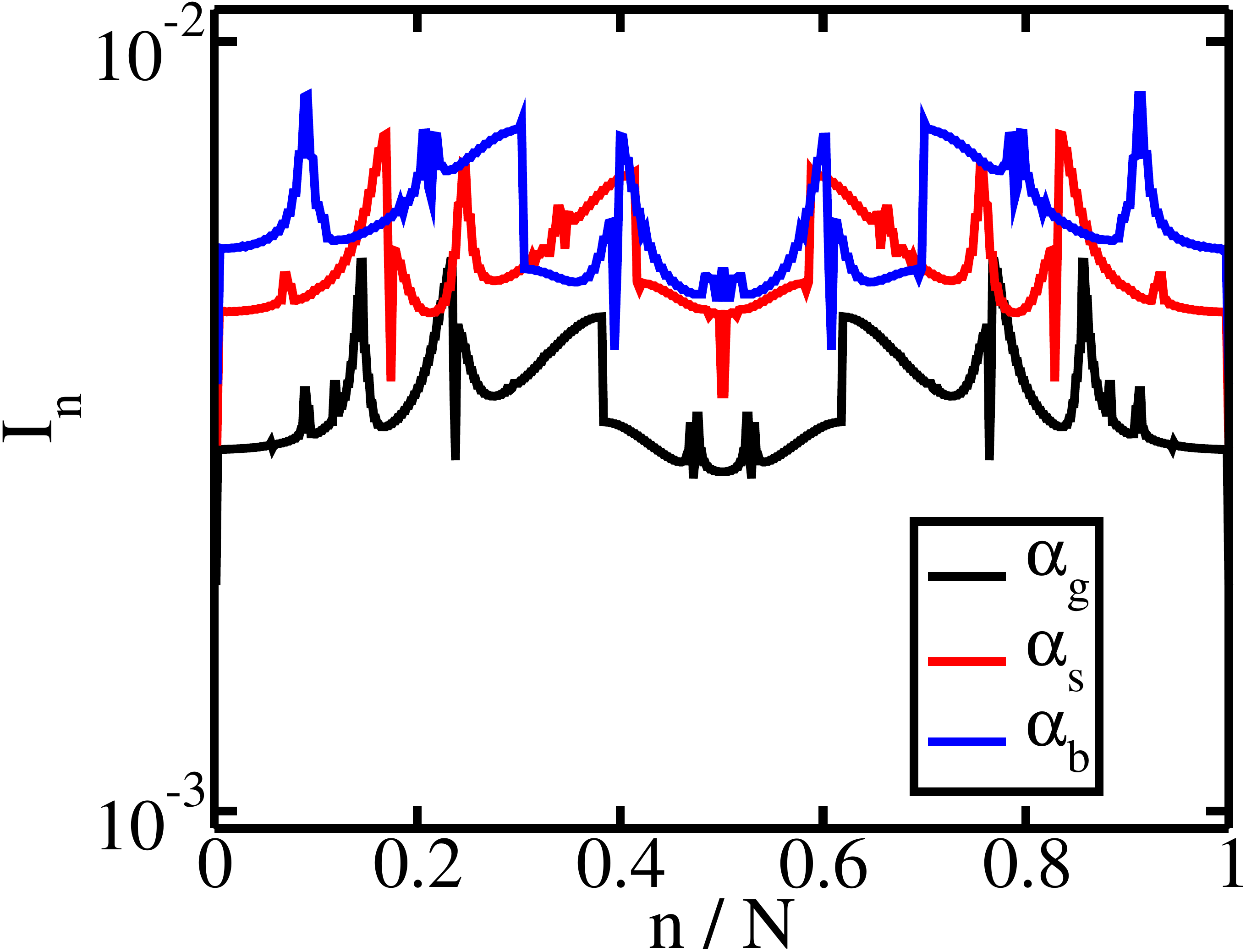}}{(b)}\\
		  	\stackunder{\includegraphics[width=4.2cm,height=3.65cm]{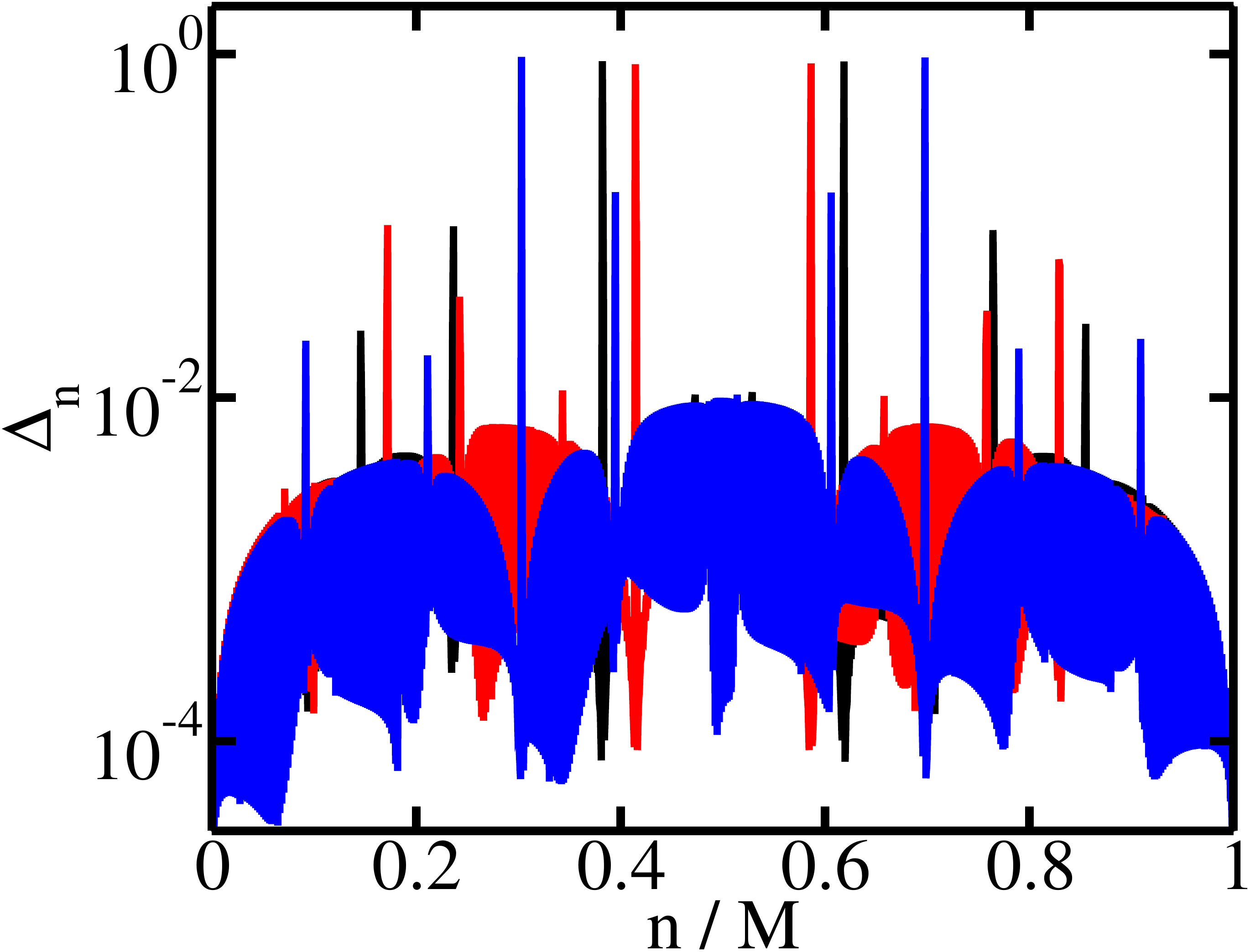}}{(c)}\hspace{0.15cm}
		  	\stackunder{\includegraphics[width=4.2cm,height=3.65cm]{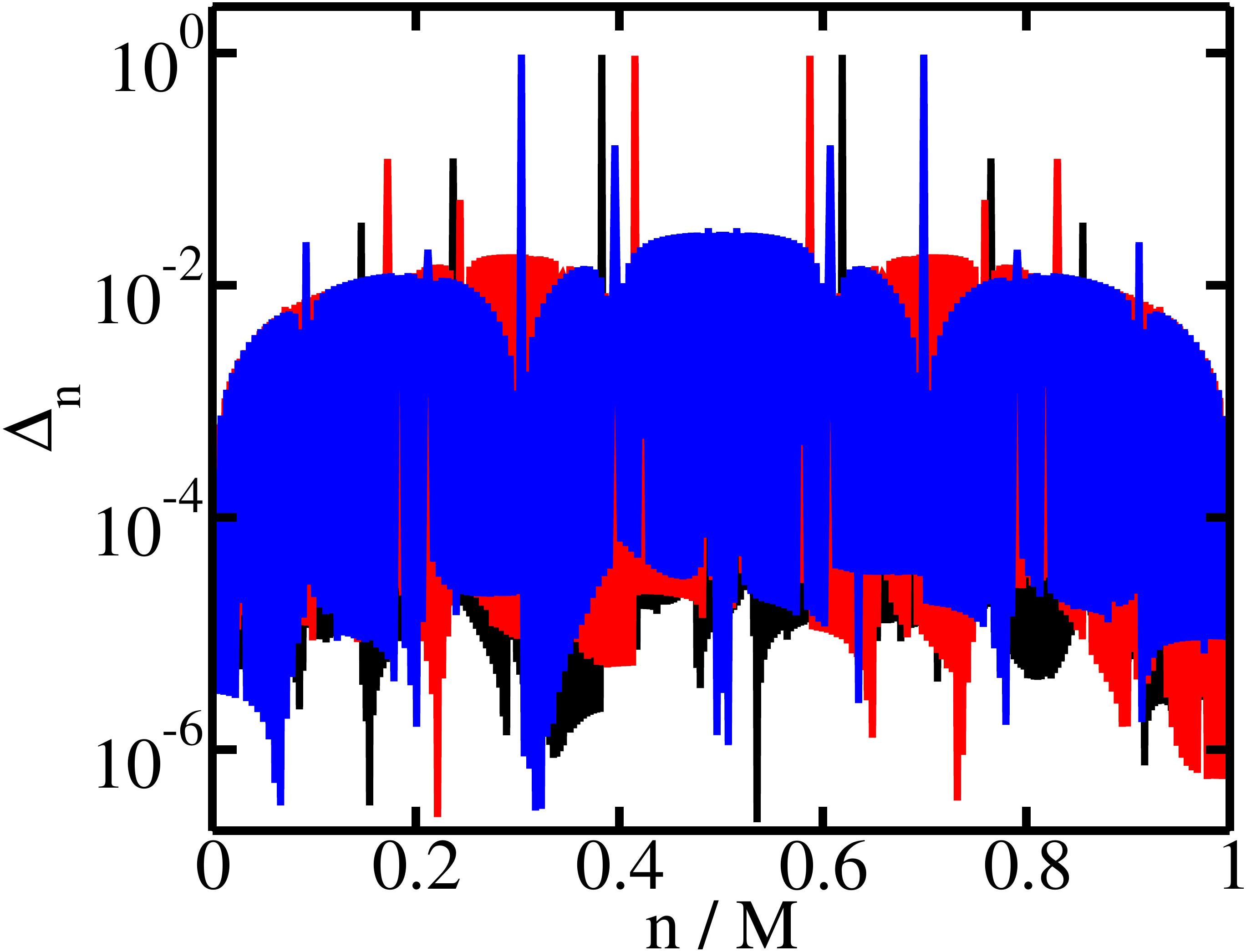}}{(d)}
		  	\caption{(a) $IPR$ of the single particle eigenstates $I_n$ for different values of $\alpha$ and fixed $N=1024$. (b) Similar plots for $N=610,408$ and $360$ for $\alpha_g,\alpha_s$ and $\alpha_b$ respectively. For these plots $n/N$ in the x-axis stands for the fractional index of eigenstates. (c) Consecutive level-spacings $\Delta_n=E_{n+1}-E_n$ for different values of $\alpha$ and fixed $N=1024$. (b) $\Delta_n$'s for $N=610,408$ and $360$ for $\alpha_g,\alpha_s$ and $\alpha_b$ respectively. $n/M$ in the x-axis stands for the fractional index of level-spacings, where total number of spacings $M=N-1$. For all the plots $\lambda=1$ in the AAH model. The legend shown in figure (b) applies also to figures (a), (c) and (d).} 
	\label{ipr_aah}
\end{figure}

{\it IPR}: Inverse participation ratio ($IPR$) of all the single particle eigenstates
for $\lambda=1$ (delocalized phase) is shown in Fig.~\ref{ipr_aah}(a)
for a non-Fibonacci $N=1024$ and different values of $\alpha$. There
exist eigenstates with high $IPR$ for fractional index $n/N=\alpha_g,
\alpha_g^2,\alpha_g^3$ $(\approx0.618,0.382,0.236)$ etc. for the
golden mean. Similarly high-$IPR$ eigenstates are also found for the
cases of silver mean ($\alpha_s$) and bronze mean ($\alpha_b$) at
$n/N=\alpha_s+\alpha_s^2,\alpha_s,\alpha_s^2 + \alpha_s^3,
\alpha_s^2,...$ $(\approx0.58,0.41,0.24,0.17,...)$ etc. and
$n/N=2\alpha_b+\alpha_b^2,\alpha_b+\alpha_b^2,\alpha_b,...$
$(\approx0.69,0.39,0.3,...)$ respectively. The single-particle energy
spectra of these systems show large gaps at the positions where the
high-$IPR$ states exist~\cite{roy2019study} as shown in
Fig.~\ref{ipr_aah}(c). In this figure the level-spacing $\Delta_n =
E_{n+1}-E_n$ with $E_n$ being the energy of the
$n\textsuperscript{th}$ eigenstate. Total number of level-spacings
$M=N-1$.

\begin{figure}
\centering
	\stackunder{\includegraphics[width=4.2cm,height=3.9cm]{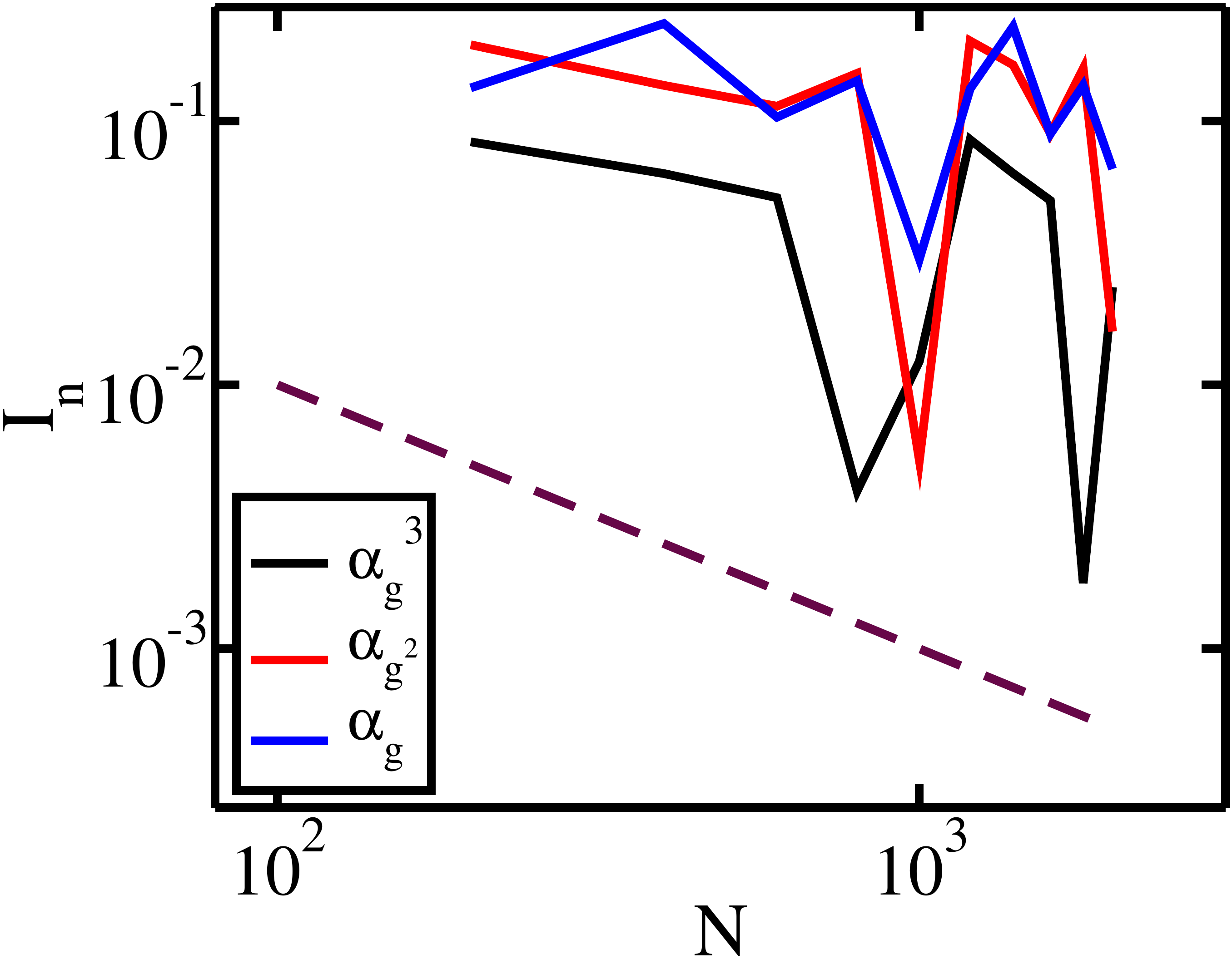}}{(a)}\hspace{0.15cm}
	\stackunder{\includegraphics[width=4.2cm,height=3.9cm]{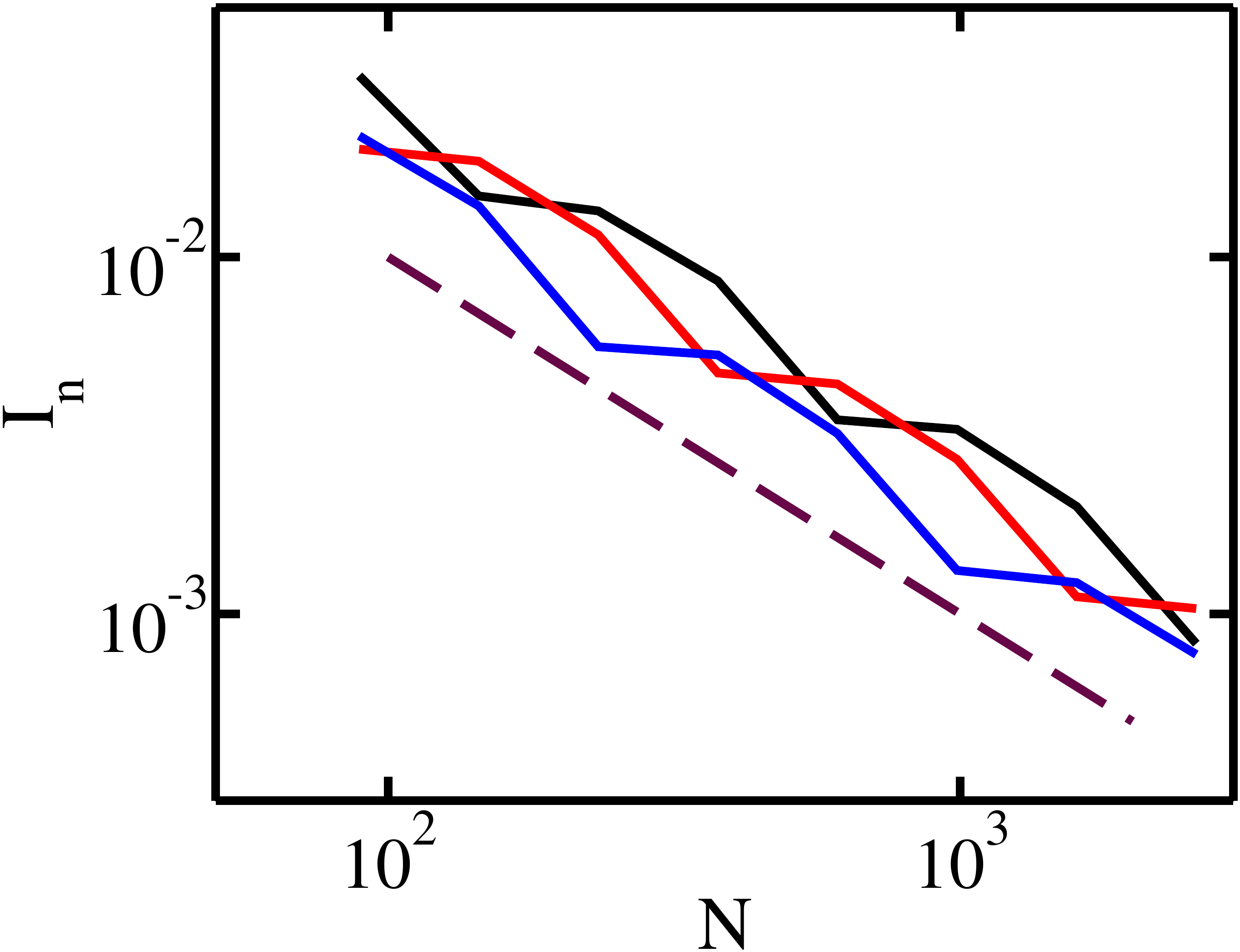}}{(b)}
	\caption{(a) $IPR$ of the special eigenstates with fractional index $n/N=\alpha_g^3,\alpha_g^2,\alpha_g$ as a function of system size $N$, which is a non-Fibonacci number. (b) Similar plots for $N$, which is a Fibonacci number corresponding to $\alpha_g$. For all the plots $\lambda=1$. The dashed line represents $1/N$ dependence of $IPR$ of the non-special delocalized eigenstates.}
	\label{iprscaling_aah}
\end{figure}

These high-$IPR$ eigenstates seem to vanish if $N$ is chosen to be a
Fibonacci number as shown in Fig.~\ref{ipr_aah}(b) for $\lambda=1$ and
$N=610,360$ and $408$ for $\alpha_g,\alpha_s$ and $\alpha_b$
respectively. However, we remark that the large gaps still continue to
persist in the energy spectra as also shown in Fig.~\ref{ipr_aah}(d).
The high-$IPR$ eigenstates show an anomalous system size
dependence. As an example we show the scaling of $IPR$ of the special
eigenstates with $N$ in Fig.~\ref{iprscaling_aah} for $\lambda=1.0$
and $\alpha_g$. Here $N$ is restricted respectively to be
non-Fibonacci and Fibonacci in Fig.~\ref{iprscaling_aah}(a) and
(b). For non-Fibonacci $N$ the scaling behavior is severely anomalous
and deviates from $1/N$. For Fibonacci $N$ the scaling behavior is
less anomalous and close to $1/N$ (although not exactly $1/N$) which is
represented by the dashed line for non-special delocalized
eigenstates.
\\

\begin{figure}
	\centering
	        \stackunder{\includegraphics[width=4.0cm,height=3.9cm]{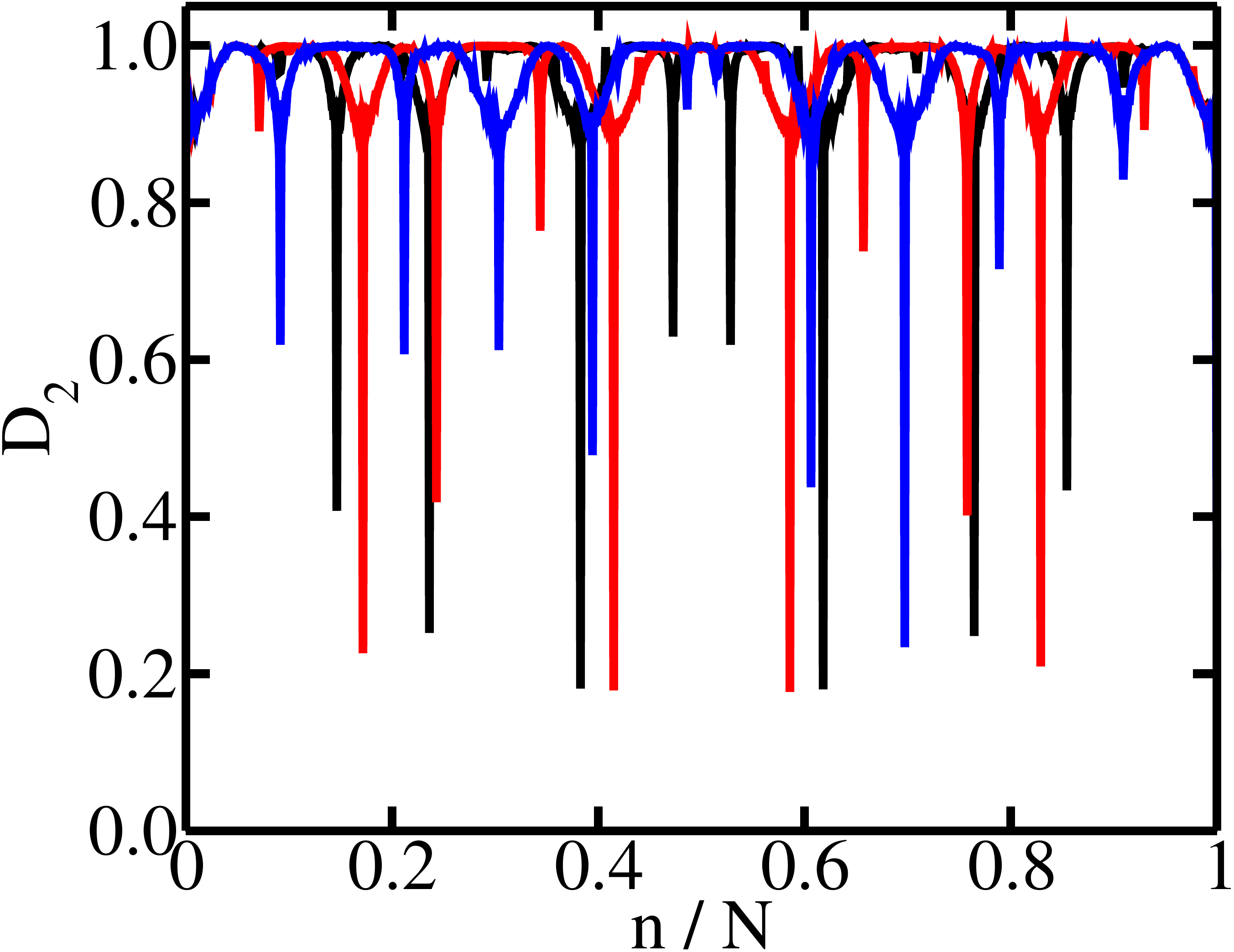}}{(a)}\hspace{0.15cm}
	        \stackunder{\includegraphics[width=4.0cm,height=3.9cm]{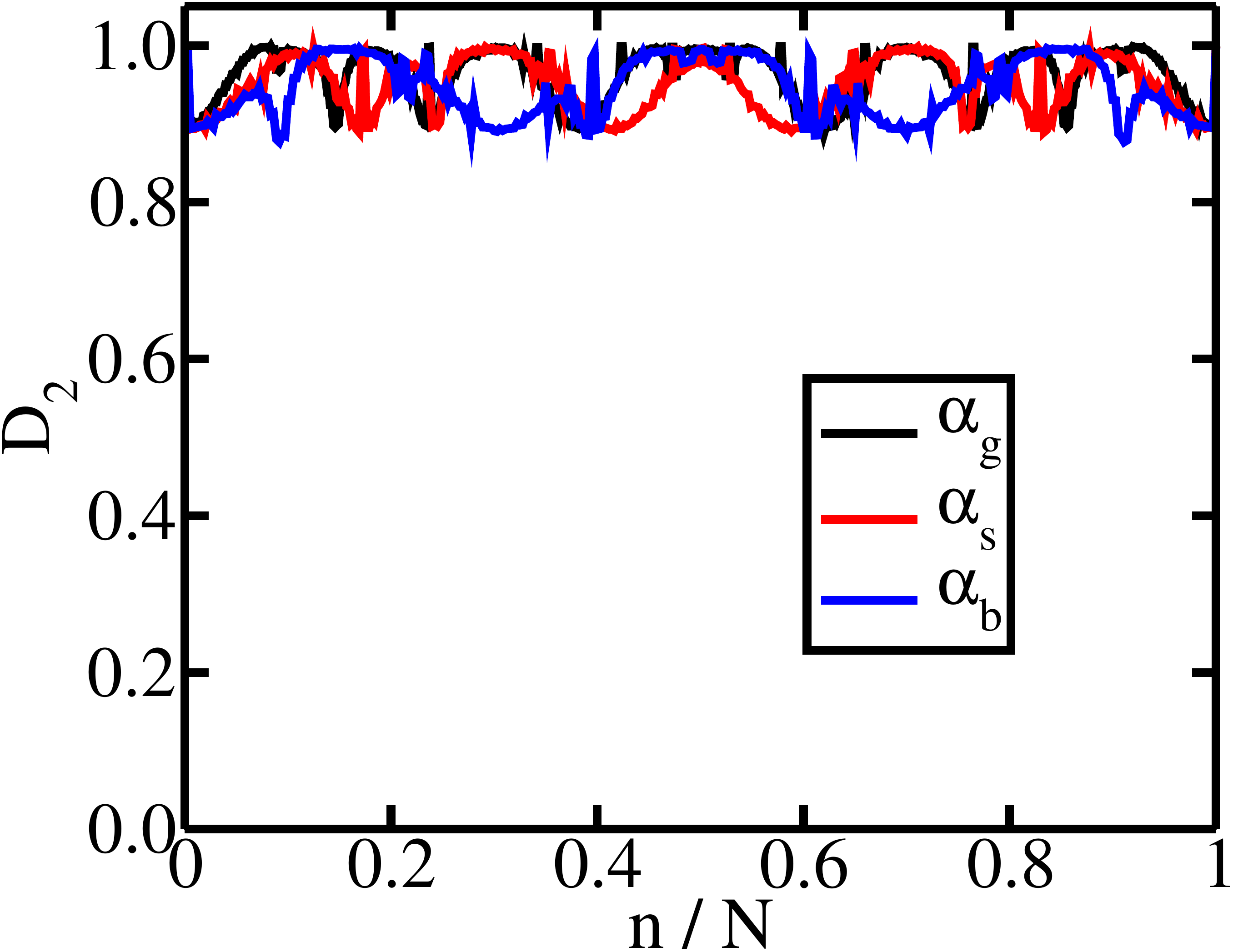}}{(b)}
	        \caption{(a) Fractal dimension $D_2$ of the single
                  particle eigenstates for different values of
                  $\alpha$ and fixed $N=1000$. (b) Similar plots for
                  $N=610,408$ and $360$ for $\alpha_g,\alpha_s$ and
                  $\alpha_b$ respectively. For all the plots
                  $\lambda=1$ in the AAH model. $n/N$ in the x-axis
                  stands for fractional index. Here
                  $\delta=1/{N_l}=0.01$.}
	\label{D2_aah}
\end{figure}

{\it Fractal dimension}: 
The fractal dimension $D_2$ is calculated for each single particle eigenstate for $\lambda=1$ and different parameters
$\alpha_g, \alpha_s$ and $\alpha_b$ in a system of non-Fibonacci
number of sites $N=1000$ as shown in Fig.~\ref{D2_aah}(a).
In the delocalized phase $D_2\approx1$ for the
majority of the eigenstates. 
The large deviations from $D_2\approx1$ are observed at the fractional eigenstate index
$n/N\approx\alpha_g, \alpha_g^2,\alpha_g^3$ etc. for $\alpha_g$. Similar deviations can be seen at
$n/N\approx\alpha_s+\alpha_s^2,\alpha_s,\alpha_s^2 + \alpha_s^3,
\alpha_s^2$ etc. for $\alpha_s$, and
$n/N\approx2\alpha_b+\alpha_b^2,\alpha_b+\alpha_b^2,\alpha_b$ etc. for
$\alpha_b$. For these special eigenstates $0<D_2<1$ which implies the presence of
non-delocalized states. Fig.~\ref{D2_aah}(b) indicates that the large
fluctuations of $D_2$ seem to vanish and $D_2$ is close to $1$ for all the eigenstates when a
Fibonacci number is chosen for
$N$. This can be understood from Fig.~\ref{iprscaling_aah}(b).
\begin{figure}
	\centering
	        \includegraphics[width=6.3cm,height=4.8cm]{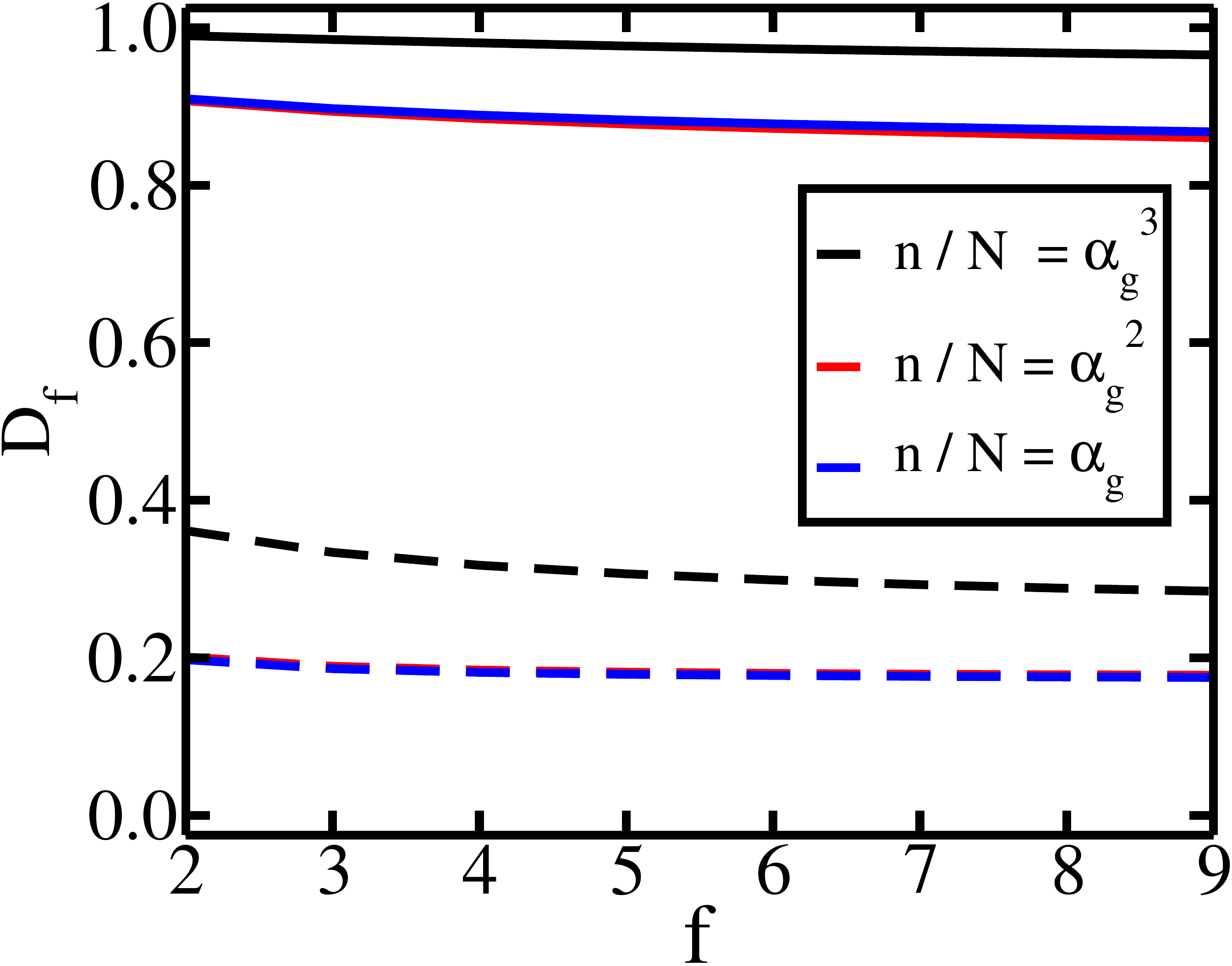}
		  	\caption{Fractal dimension $D_f$ as a function of $f$ for the single particle eigenstates with fractional index $n/N=\alpha_g^3,\alpha_g^2,\alpha_g$. The solid lines represent plots for Fibonacci $N=610$ whereas the dashed lines represent plots for non-Fibonacci $N=1000$. For all the plots, $\lambda=1$, $\alpha=\alpha_g$ and $\delta=1/{N_l}=0.01$.} 
	\label{Dq_aah}
\end{figure}

In Fig.~\ref{Dq_aah} we show the fractal dimension $D_f$ as a function
of $f$ for the eigenstates with fractional index
$n/N=\alpha_g^3,\alpha_g^2,\alpha_g$ for $\lambda=1$ and golden mean 
$\alpha_g$. In this figure the solid lines represent the plots for
Fibonacci $N=610$ whereas the dashed lines represent the plots for
non-Fibonacci $N=1000$. We observe that the solid lines change very
little with $f$ and are close to $1$. Here $D_f$ deviates a little
from $1$ because these eigenstates are not perferctly delocalized as
depicted in Fig.~\ref{iprscaling_aah}(b). On the other hand the dashed
lines show a small variation with $f$ and their typical value is just
a fraction of one. This indicates that for non-Fibonacci $N$, the
special eigenstates with high $IPR$ are weakly multifractal whereas
for Fibonacci $N$, the special eigenstates behave more like the (imperfect) delocalized states like all other non-special states. This is true even
for silver mean $\alpha_s$ and bronze mean $\alpha_b$ (not shown
here). 
\\

\begin{figure}
	\centering
	        \stackunder{\includegraphics[width=4.0cm,height=3.9cm]{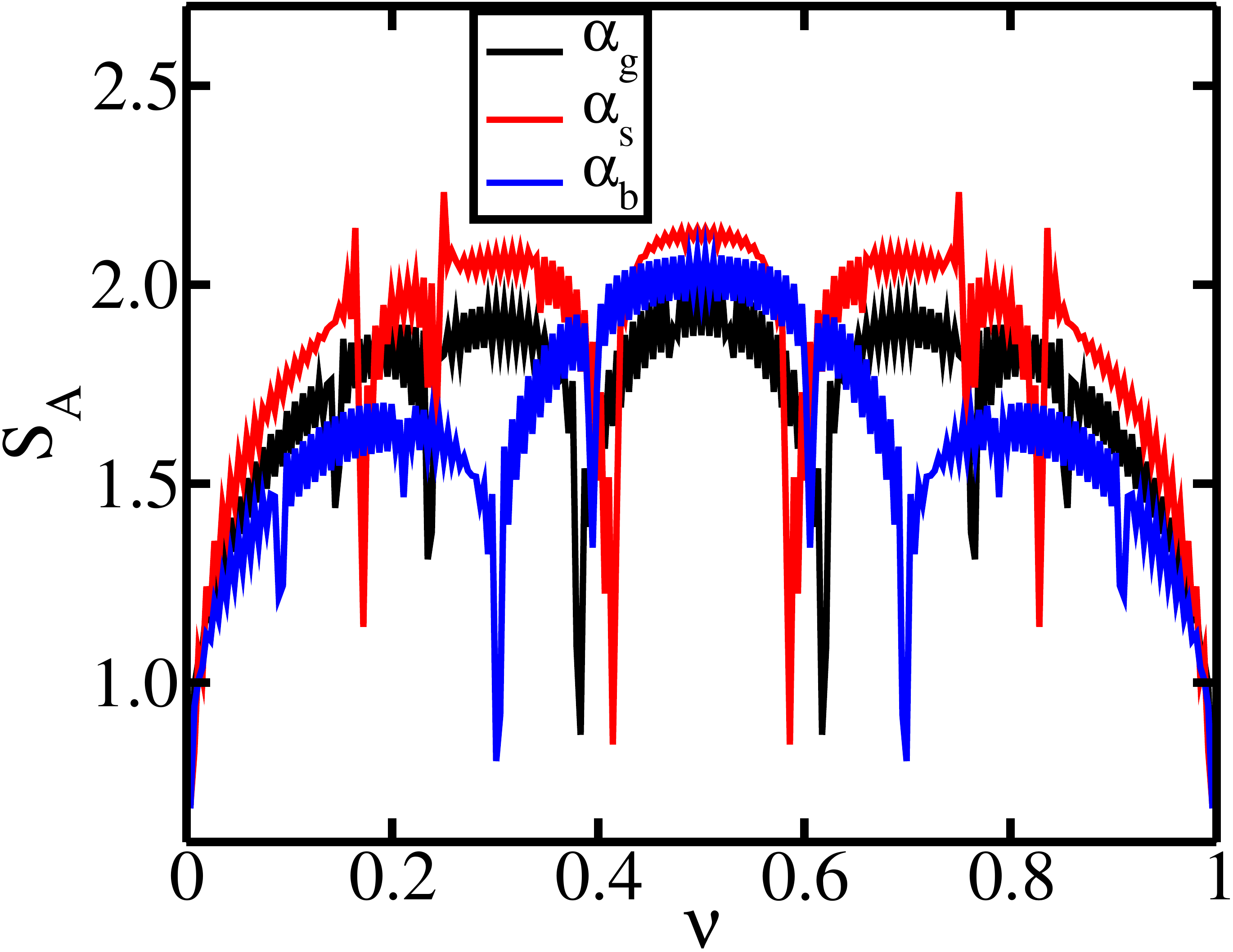}}{(a)}\hspace{0.15cm}
		  	\stackunder{\includegraphics[width=4.0cm,height=3.9cm]{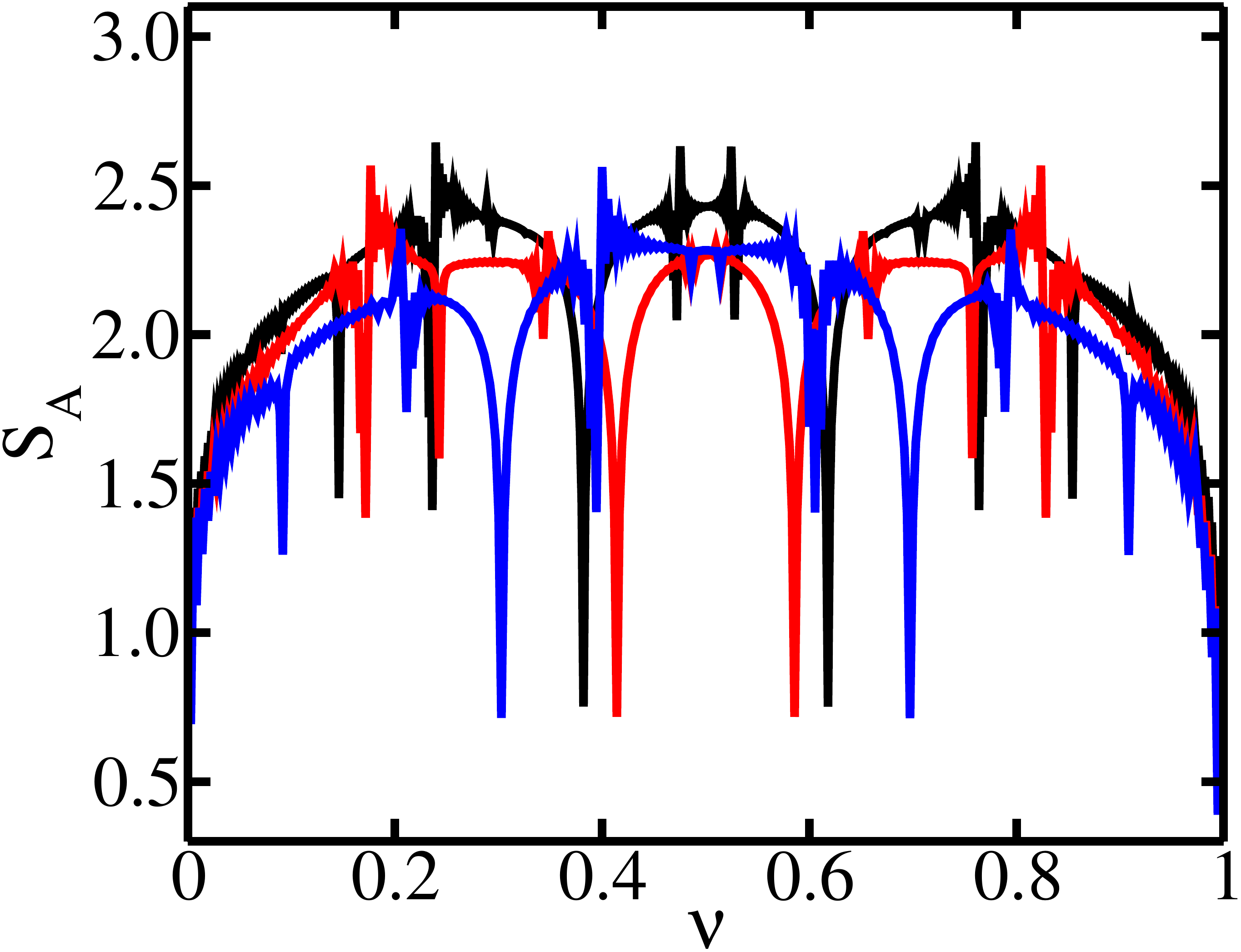}}{(b)}
		  	\caption{(a) Entanglement entropy $S_A$ of the ground state as a function of fermionic filling $\nu$ for different values of $\alpha$ and fixed $N=256$. (b) Similar plots for $N=610,408$ and $360$ for $\alpha_g,\alpha_s$ and $\alpha_b$ respectively. For all the plots $\lambda=1$ in the AAH model and size of subsystem A is $L=N/2$.} 
	\label{ee_aah}
\end{figure}

{\it Entanglement entropy}: The ground state entanglement entropy
$S_A$ of half the system (subsystem $L=N/2$) as a function of filling
fraction $\nu$ for $\lambda=1$ is shown in Fig.~\ref{ee_aah}(a) for a
non-Fibonacci $N=256$ and different values of $\alpha$. Here
$\nu=N_p/N$ where $N_p$ and $N$ are the number of particles and number
of sites respectively. Similar to high $IPR$ in Fig.~\ref{ipr_aah}(a),
significantly low $S_A$ is found at $\nu\approx\alpha_g,
\alpha_g^2,\alpha_g^3$ etc. for $\alpha_g$;
$\nu\approx\alpha_s+\alpha_s^2,\alpha_s,\alpha_s^2 + \alpha_s^3,
\alpha_s^2$ etc. for $\alpha_s$;
$\nu\approx\alpha_b+\alpha_b^2,\alpha_b+\alpha_b^2,\alpha_b$ etc. for
$\alpha_b$.  But in contrast to Fig.~\ref{ipr_aah}(b) of $IPR$, the
low $S_A$ regions seem to persist as shown in Fig.~\ref{ee_aah}(b)
even for Fibonacci $N=610,408,360$ for $\alpha_g,\alpha_s,\alpha_b$
respectively. The persisting imperfection of the special eigenstates (shown in Fig.~\ref{iprscaling_aah}) may be a reason behind this. The imperfection is captured at a magnified level by the many-particle entanglement entropy as compared to single particle $IPR$ for a Fibonacci $N$.       
In the delocalized phase, $S_A\propto\ln
L$~\cite{roy2019study} for all values of $\nu$ except for the special
values of $\nu$ where $S_A$ abides by the `area law' with
significantly smaller magnitudes. The signature of criticality in the
model is absent for special $\nu$. These properties of the special
$\nu$ have been shown earlier in Ref.~\onlinecite{roy2019study} for
$\alpha_g$ and hold good for $\alpha_s$ and $\alpha_b$ also. However,
the non-special half-filled ($\nu=0.5$) ground state shows $S_A\propto
\ln L$ both in the delocalized phase and at the critical point (almost
$\ln L$)~\cite{prb,roosz2020entanglement} whereas $S_A\propto L^0$ in the localized
phase with the prefactor $K$ of the logarithmic term being approximately $0.33,0.26$ and $0$ respectively. The logarithmic scaling at the critical point shows that the multifractal states are extended in nature but nonergodic as the prefactor differs from the ergodic delocalized ones. 

\end{document}